\documentclass[pra,aps,twocolumn, superscriptaddress, nofootinbib, tightenlines, nobibnotes]{revtex4-1}

\usepackage{amssymb}
\usepackage{graphicx}
\usepackage{dcolumn}
\usepackage{gensymb}
\usepackage{multibib}
\usepackage{bm}
\usepackage{hyperref}
\usepackage{tikz}
\usepackage{comment}
\usepackage{xcolor}
\usepackage[caption=false]{subfig}
  \usepackage[utf8]{inputenc}
 \usepackage[english]{babel}
\usepackage{amsmath}
 \usepackage{hyperref}
 \usepackage{amsmath}
 \usepackage{mathtools} 
 \usepackage{braket}

\definecolor{lapislazuli}{rgb}{0, 0, 1}
\definecolor{YKblue}{rgb}{0.0, 0.18, 0.65}
\definecolor{carmine}{rgb}{0.81, 0.09, 0.03}
\definecolor{lavender}{rgb}{0.84, 0.49, 0.87}

\hypersetup{
colorlinks=true,
linkcolor=lapislazuli,
linktoc=page,
citecolor=lapislazuli,
urlcolor=lapislazuli
}

\usepackage{amssymb}
\usepackage{mathtools}

\everymath{\displaystyle} 
\newcommand{\pr}[1]{\ensuremath{\left[#1\right]}} 
\newcommand{\pc}[1]{\ensuremath{\left(#1\right)}} 
\newcommand{\av}[1]{\ensuremath{\left\langle#1\right\rangle}} 

\begin{document}

\title{Generating Einstein–Podolsky–Rosen correlations for  teleporting collective spin states in a two dimensional trapped ion crystal}

\author{Muhammad Miskeen Khan}
	\affiliation{JILA, National Institute of Standards and Technology and University of Colorado, 440 UCB, Boulder, Colorado 80309, USA}
	\affiliation{Center for Theory of Quantum Matter, University of Colorado, Boulder, Colorado 80309, USA}
	
	
	\author{Edwin Chaparro}
	\affiliation{JILA, National Institute of Standards and Technology and University of Colorado, 440 UCB, Boulder, Colorado 80309, USA}
	\affiliation{Center for Theory of Quantum Matter, University of Colorado, Boulder, Colorado 80309, USA}

 \author{Bhuvanesh Sundar$^*$}
	\affiliation{JILA, National Institute of Standards and Technology and University of Colorado, 440 UCB, Boulder, Colorado 80309, USA}
        \thanks{Now at Rigetti Computing, 775 Heinz Avenue, Berkeley, California 94710, USA}

	\author{Allison Carter}
	\affiliation{National Institute of Standards and Technology, Boulder, Colorado 80309, USA}

	\author{John Bollinger}
	\affiliation{National Institute of Standards and Technology, Boulder, Colorado 80309, USA}

 \author{Klaus Molmer}
	\affiliation{Niels Bohr Institute, University of Copenhagen, 2100 Copenhagen, Denmark}

	\author{Ana Maria Rey}
\affiliation{JILA, National Institute of Standards and Technology and University of Colorado, 440 UCB, Boulder, Colorado 80309, USA}
	\affiliation{Center for Theory of Quantum Matter, University of Colorado, Boulder, Colorado 80309, USA}
\begin{abstract}
We propose the use of phonon-mediated interactions as an entanglement resource to engineer Einstein–Podolsky–Rosen (EPR) correlations and to perform teleportation of collective spin states in  two-dimensional ion crystals.  We emulate  continuous variable quantum teleportation protocols between subsystems corresponding  to different nuclear spin degrees of freedom.  In each of them,  a quantum state  is encoded in an electronic spin degree of freedom that couples to the vibrational modes of the crystal.  We show that  high fidelity teleportation of  spin-coherent states and their phase-displaced variant, entangled spin-squeezed states, and  Dicke states, is possible  for realistic
experimental conditions in arrays from a few tens to a few hundred ions.
\end{abstract}
\maketitle                  

\textit{Introduction:}\label{sec:Introduction}
Correlated quantum states, such as entangled spin-squeezed states, have been predicted to offer a significant gain in sensing and communication applications \cite{RevModPhys.90.035005}. While great progress  has been achieved using macroscopic atomic ensembles in optical cavities or vapor cells \cite{Hosten2016, Cox2016,Appel2009,Bao2020,Colombo2022,     Robinson2024,Bornet2023}, these systems typically lack the level of  quantum control  over motional  degrees of freedom desired for more general quantum information tasks. Arrays of two-dimensional trapped-ion crystals \cite{ Britton2012,Bohnet2016,Grttner2017,Gilmore2021,Mavadia2013,PRXQuantum.4.020317} are  emerging  as a promising platform 
where one can scale up the number of ions  while retaining full or partial quantum control over vibrational modes and all-to-all internal state connectivity. These capabilities can thus open an exciting  opportunity
for entanglement generation.  
In the context of quantum information processing, quantum teleportation is  one of the most useful resources \cite{PhysRevLett.70.1895, PhysRevA.49.1473, PhysRevLett.80.869, Pirandola2015, Hu2023}  that  unravels the power of  entanglement  for  quantum communication and information processing tasks \cite{Gottesman1999, Barrett2004, Olmschenk2009,Gao2010,Chou2018, Riebe2004, PhysRevX.12.031013,PRXQuantum.4.010321,Rugg2019}. While teleportation of optical and spin coherent states have been experimentally demonstrated using large atomic ensembles \cite{Krauter2013,Sherson2006}, teleportation of collective entangled spin states in trapped ions platforms that enjoy control over both internal and external degrees of freedom \cite{RevModPhys.93.025001,Wan2019,Landsman2019}, as required for most  quantum information processing tasks, is still pending.%
%

\begin{figure}[t!]
	\includegraphics[width=1\columnwidth]{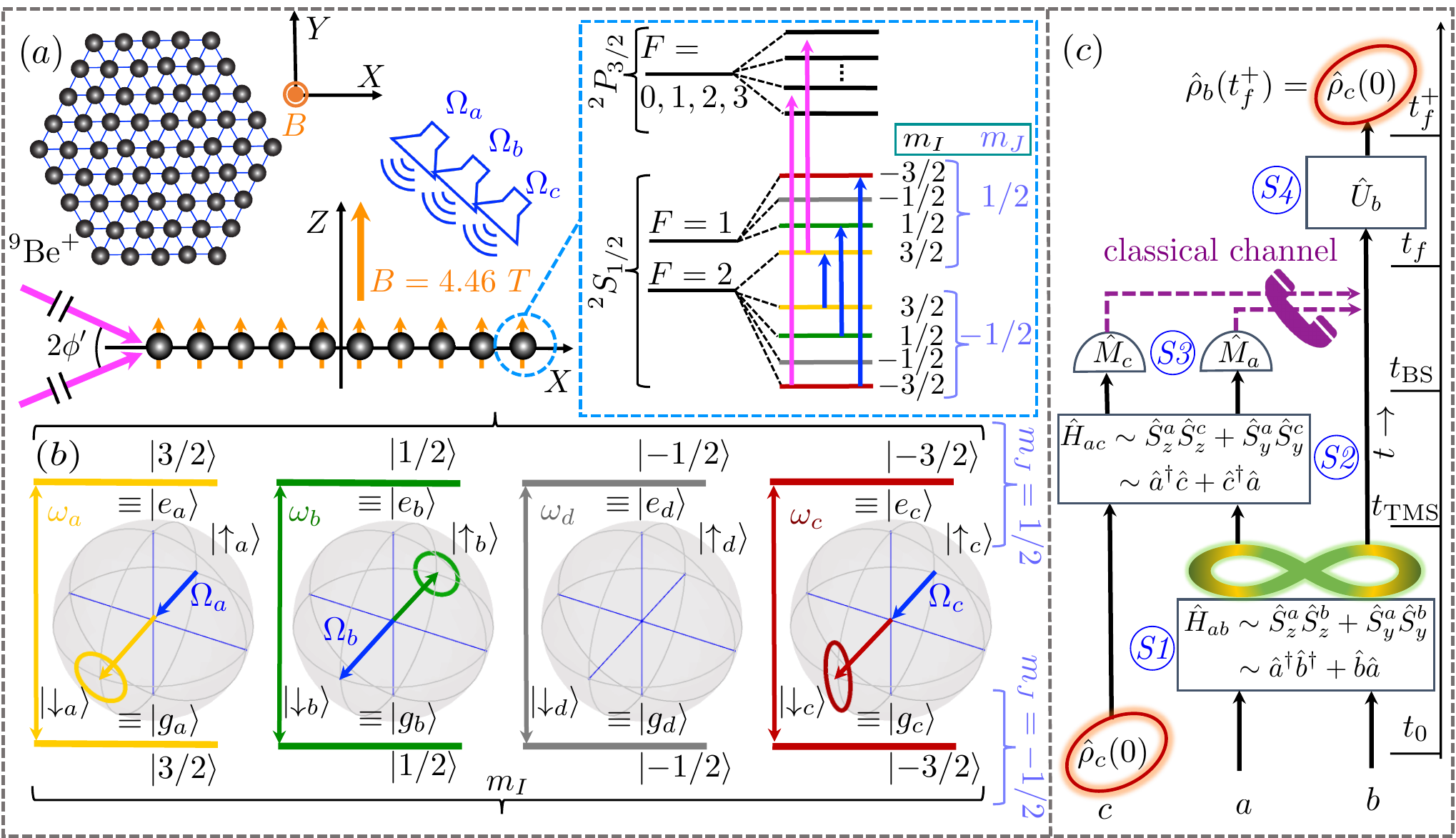}
	\caption{(a): Schematic for a two-dimensional trapped ion crystal made of  $^9 \rm Be^{+}$ ions located in the  $(X,Y)$-plane as realized in the Penning trap at NIST \cite{Bollinger2013}. A strong magnetic field $B$ results in  splitting of the internal Zeeman levels as shown in the inset. The nuclear spins form distinctly addressable ensembles  $l=a,b,c$. An optical dipole force (ODF) implemented by Raman beams [pink] couples each nuclear spin ensemble to a common transverse vibrational mode of the ions. Three spin ensembles are resonantly driven by microwaves with Rabi frequencies $\Omega_{l}, ~ l=\{a,b,c\}$, to implement the teleportation circuit. (b) The initial states of the nuclear spin ensemble are shown on the corresponding Bloch spheres. From left to right,  spin ensemble $a$, $b$, and $c$ are initialized as  a spin-polarized state along $-x$ [yellow], $+x$ [green], and $-x$ [red] directions, respectively. (c) Schematics of the teleportation protocol: it consists of different stages ({\it S1-S4}) as discussed in the text .}
\label{fig:telepfig1}
\end{figure}
Here we demonstrate that  quantum teleportation of collective  spin-states can be implemented in current 2D crystal arrays in a Penning trap using the center-of-mass motional mode as an entanglement resource. The proposed scheme is analogous to the continuous variable quantum teleportation scheme of Braunstein and Kimble (BK) \cite{PhysRevLett.80.869}, however, instead of relying on measurement based schemes for entanglement generation \cite{PhysRevLett.85.5643}, our system uses  unitary phonon-mediated all-to-all spin-spin interactions. Measurements, enabled by spectroscopic resolution of internal spin levels of ions,  are only used  for the act of teleportation \cite{Sherson2006} at the end of the protocol. The access to long-range spin-spin interactions in trapped ions arrays \cite{Britton2012, RevModPhys.93.025001}, allow for  
 the initialization of entangled states \cite{Bohnet2016},  which we show can also be teleported. Our protocol  thus  opens  a path for the implementation of continuous variable quantum information protocols \cite{RevModPhys.77.513} without requiring additional non-linear interactions that are typically absent in pure photonic and phononic systems. Moreover, since the phonon modes in  trapped-ion systems are very long-lived, our system   does not suffer from  the detrimental losses faced by photons.Finally, while  we focus here on states within the same spatial ensemble, the applicability of the same protocols in  bilayer arrays or 3D crystals \cite{hawaldar2024bilayer} could enable  the possibility to  teleport states between  spatially separated  layers.

\textit{Setup:}\label{sec:Setup}
We consider a two-dimensional trapped ion crystal made of $^9 \rm Be^{+}$ ions in a Penning trap as schematically shown in Fig.~\ref{fig:telepfig1} (a). The crystal is located in the $X,Y$-plane and it is subjected to a strong magnetic field ($B\sim 4.5~{\rm T}$)  along  the $Z-{\rm direction}$ that sets the quantization axis and allows us to work in the Paschen-Back regime \cite{PB2009} with  decoupled  electronic (with $J=1/2$) and nuclear (with $I=3/2$) hyperfine-Zeeman states as shown in the inset of the Fig.~\ref{fig:telepfig1} (a). In our scheme,  the  ions are initialized  in three out of a total four nuclear spin states \cite{PhysRevA.84.012510}, $m_{I}=3/2(l=a), 1/2(l=b),-3/2(l=c)$, with  $N_{l=a,b,c}$ ions in each nuclear spin ensemble. In this way we have access to three distinct choices of qubit degrees of freedom  per atom,  each one characterized by the 
$\{\ket{e_{l}},\ket{g_{l}}\}$ levels with energy splitting $\omega_{l}$, where  the labels $g,e$ denote the $m_{J}=-1/2,1/2$ electronic states respectively. Thanks to the   large energy separation,  $\omega_{ll^{\prime}}\equiv\omega_{l}-\omega_{l^{\prime}} \sim 500~ {\rm MHz}$ \cite{PhysRevA.84.012510},   the different nuclear spin sub-ensembles can be independently  controlled  with negligible  coupling between them by  microwave drives, with Rabi drive strength $\Omega_{l}$ and frequency $\omega_{l}^\mathcal{D}$, as schematically shown in Fig.~\ref{fig:telepfig1} (b). 

We further assume the ions are driven by interfering laser beams with a beat-note frequency $\mu$ as shown in Fig.~\ref{fig:telepfig1} (a). The beams  are applied off-resonantly (detuned by $\sim$20 GHz) to  the nearest  optical transition spanned by the $^2P_{3/2}$ manifold. Their  polarization and  orientation  are set to couple the spin degree of freedom to the axial modes of the crystal and $\mu$   is set  to be close to the   center-of-mass (COM) mode frequency  of the crystal, $\omega_M$ to avoid excitation of other modes (cf. inset of Fig.~\ref{fig:telepfig1} (a)).
The net result is the generation of an  electronic-spin-dependent optical dipole force (ODF) on  the ions \cite{Britton2012} that  acts approximately in the opposite direction for the $m_J=\pm 1/2$ states (up to a small correction $\epsilon_l$ which turns out to be irrelevant for the physics in consideration (see \cite{SupplementalMaterials})).

Assuming the ions have an axial extent that is small compared to the wavelength of the moving lattice, by going to the rotating frame of the beat-note frequency, the Hamiltonian of the total system  can be written as $(\hbar=1)$:
\begin{equation}\textstyle{\hat{H}=\hat{H}_{s}+\hat{H}_{\rm ODF}}, \label{eq:Hamiltonian4}\end{equation} where  $\textstyle{ \hat{H}_{s}={\sum}_{l=a,b,c}\pr{\omega_{l}\hat{S}_{z}^{l}+\Omega_{l}/2 \pc{\hat{S}_{+}^{l}e^{-i\omega_{l}^{\mathcal{D}}t}+{\rm H.C.}}} }$ describes the applied  microwave drives and $\textstyle{ \hat{H}_{\rm ODF}
= \delta_{M}\hat{m}^{\dagger}\hat{m}+{\sum}_{l=a,b,c}\frac{g_{l}}{\sqrt{N}}(\hat{m}+\hat{m}^{\dagger}) \hat{S}_{z}^{l} }$
 the ODF Hamiltonian with $\delta_{M}=\omega_{M}-\mu$ \cite{PhysRevLett.121.040503}. Here, we introduced  the COM phonon annihilation (creation) operator $\hat{m} (\hat{m}^{\dagger})$  and $\textstyle{ S_{\alpha}^{l}\equiv\frac{1}{2}\sum_{j=1}^{N_{l}}\sigma_{\alpha}^{l,j} }$ the collective spin operators, with $\alpha=\pc{x,y,z}$  and $\hat{S}_{\pm}^{l}\equiv \hat{S}_{x}^{l} \pm i \hat{S}_{y}^{l}$ the corresponding raising and lowering operators. The spin-phonon coupling is denoted by $g_{l}$, and  $\textstyle {N=\sum_{l}N_{l}}$ is the total number of ions in crystal.  Note that the values $\textstyle{N_{l}}$ are set by the initial preparation and they are conserved during the interaction of the ions. For each ensemble, the single ion coupling $g_{l}$ is inversely dependent to the one-photon Raman detuning, and thus can be slightly different for each of the nuclear spin ensembles by the order of just a few percent.
 
In a frame rotating at the microwave drive frequency, resonant with the nuclear spin transition $\omega_{l}=\omega_{l}^{\mathcal{D}}$, we rewrite the Hamiltonian $\hat{H}$ in the dressed (rotated) basis. The dressed states are eigenstates of the microwave drives,  and are explicitly given by $\ket{\uparrow_{l}}=(|g_{l}\rangle+|e_{l}\rangle)/\sqrt{2}$ and $\ket{\downarrow_{l}}=(|g_{l}\rangle-|e_{l}\rangle)/\sqrt{2}$ as shown in Fig.~\ref{fig:telepfig1} (b) on the corresponding Bloch spheres. In terms of the collective spin operators in the dressed frame ($-\mathcal{\hat{S}}^{l}_{x}\rightarrow \hat{S}^{l}_{z}$, $\mathcal{\hat{S}}^{l}_{y}\rightarrow \hat{S}^{l}_{y} $, $\mathcal{\hat{S}}^{l}_{z}\rightarrow \hat{S}^{l}_{x}$),
the Hamiltonian reads \cite{PhysRevLett.121.040503}
\begin{align}\label{eq:FullDressedFrameHamlt}
\hat{H}=&\sum_{l}\Omega_{l}\mathcal{\hat{S}}_{z}^{l}-\sum_{l}\frac{g_{l}}{\sqrt{N}}(\hat{m}+\hat{m}^{\dagger})\mathcal{\hat{S}}_{x}^{l}+\delta_{M}\hat{m}^{\dagger}\hat{m}. 
\end{align}
We further consider the special case where  the  spin ensembles $a,~b,~c$ are initially spin-polarized  \cite{SupplementalMaterials} along $-x$, $+x$ and $-x$ direction of their corresponding  Bloch sphere, respectively  (see Fig.~\ref{fig:telepfig1} (b)). Following such initialization, the ensemble $c$ is subjected to  a unitary operation that transforms the state into a spin coherent state (which can also be slightly displaced from the initial mean magnetization), a spin squeezed state, or a Dicke state \cite{Dudin2012,PhysRevA.95.041801, Zeiher2016}. This is the state  we aim to teleport. We implement the teleportation protocol consist of different stages ({\it S1-S4}) as outlined 
in Fig. \ref{fig:telepfig1} (c), involving an entangling operation ({\it S1}) between $a$ and
$b$ ensembles, followed by a beam-splitter  ({\it S2}) interaction between $a$ and $c$ ensembles . The outcomes of the measurement performed ({\it S3}) on the latter two  are then classically communicated to ensemble $b$,
enabling us to perform spin rotations ({\it S4}) on ensemble $b$ to retrieve the teleported state .

{\it Teleportation  protocol:} For the  first stage of the teleportation protocol, we set $\Omega_{c}=0$ in Eq.~\eqref{eq:FullDressedFrameHamlt}, since we do not want  ions in this state to participate in the dynamics.  Assuming
that $|\Delta_{M}^{ab}|\equiv |\delta_M-\Omega_{ab}|\gg g_l$,
with $\Omega_{ab}\equiv(\Omega_{a}+\Omega_{b})/2$ and $\delta_{ab}\equiv(\Omega_{a}-\Omega_{b})/2$ we can adiabatically eliminate the COM phonon mode.  By going to a rotating frame defined by the unitary transformation 
$
U=e^{i\Omega_{ ab}\pc{\mathcal{\hat{S}}_{z}^{a}+\mathcal{\hat{S}}_{z}^{b}+\hat{m}^{\dagger}\hat{m}}t}$, we obtain an effective spin-spin interaction Hamiltonian of the form
\cite{SupplementalMaterials}:
\begin{align}
\hat{H}_{ab}=&
-\chi_{ab}(\mathcal{\hat{S}}_{+}^{a}\mathcal{\hat{S}}_{-}^{b}+\mathcal{\hat{S}}_{-}^{a}\mathcal{\hat{S}}_{+}^{b})\nonumber\\
&- \chi_{a a} \mathcal{\hat{S}}_{z}^{a}\mathcal{\hat{S}}_{z}^{a} - \chi_{bb} \mathcal{\hat{S}}_{z}^{b}\mathcal{\hat{S}}_{z}^{b}+ \delta_{ab}(\mathcal{\hat{S}}_{z}^{a}-\mathcal{\hat{S}}_{z}^{b}).
\label{eq:effenuclear_spin_spinHam1}
\end{align} Here   
$\chi_{\alpha \alpha'}\equiv (g_{\alpha} g_{\alpha'})/(4 \bar{N}\Delta_{M}^{ab})$, with $\{\alpha,\alpha'\}\in \{a,b\}$.
The first line in Eq.~\eqref{eq:effenuclear_spin_spinHam1} describes flip-flop processes between the two different spin ensembles.  The second line, up to  constants of motion that we have omitted,   includes the self-interaction terms plus  an  energy shift arising from the two different Rabi frequencies. 
We wish to employ this Hamiltonian to generate correlated excitations between the ensembles $a$ and $b$.  To do so, as mentioned above, we initialize the $a,b$ ensembles in fully polarized states with opposite magnetization, e.g. in  eigenstates of $\mathcal{\hat{S}}_{z}^{a}$  and $\mathcal{\hat{S}}_{z}^{b}$  with eigenvalues $-N_a/2$ and $N_b/2$, respectively. 
When $\hat{H}_{ab}$  is applied to these states, the flip-flop term 
$\mathcal{\hat{S}}_{+}^{a}\mathcal{\hat{S}}_{-}^{b}$ simultaneously generates a spin
$\ket{\uparrow_a}$ excitation in the $a$ ensemble and a spin $\ket{\downarrow_b}$ excitation in the $b$ ensemble as desired. This process, however, imposes an energy cost of $\chi_{aa}N_a + \chi_{bb}N_b$ arising from the self-interactions.  This energy penalty can be compensated by an appropriate  choice of  the Rabi frequencies  driving the $a,b$ ensembles.  Specifically  by setting $\delta_{ab}=\chi_{aa} N_a=\chi_{bb} N_b\equiv \bar{\chi}\bar{N}$  one can approximately cancel  the energy penalty \cite{SupplementalMaterials}.
The above discussion is valid in $N_{a}\sim N_{b}\gg1$ limit,
where  we  can use the mean-field approximation ( i.e. $\hat O \hat R\to \hat O \langle \hat R\rangle + \hat R \langle \hat O\rangle-  \langle \hat O\rangle  \langle \hat R\rangle$) and  approximate $\mathcal{\hat{S}}_{z}^{a}\mathcal{\hat{\mathcal{S}}}_{z}^{a} \approx 2 \langle  {\hat{\mathcal{S}}}_{z}^{a}\rangle {\hat{\mathcal{S}}}_{z}^{a}= -  N_a \mathcal{\hat{\mathcal{S}}}_{z}^{a} $ and $\mathcal{\hat{\mathcal{S}}}_{z}^{b}\mathcal{\hat{\mathcal{S}}}_{z}^{b} \approx 2 \langle  {\hat{\mathcal{S}}}_{z}^{b}\rangle {\hat{\mathcal{S}}}_{z}^{b}=   N_b \mathcal{\hat{S}}_{z}^{b}$ plus constant terms. 
\begin{figure}[t!]
	\includegraphics[width=0.472\columnwidth]{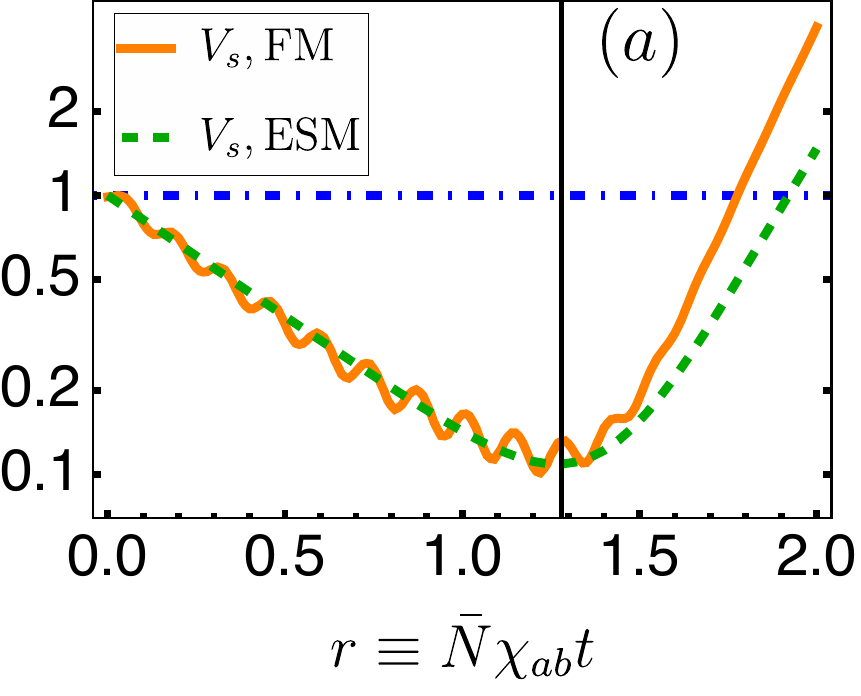}
		\includegraphics[width=0.5\columnwidth]{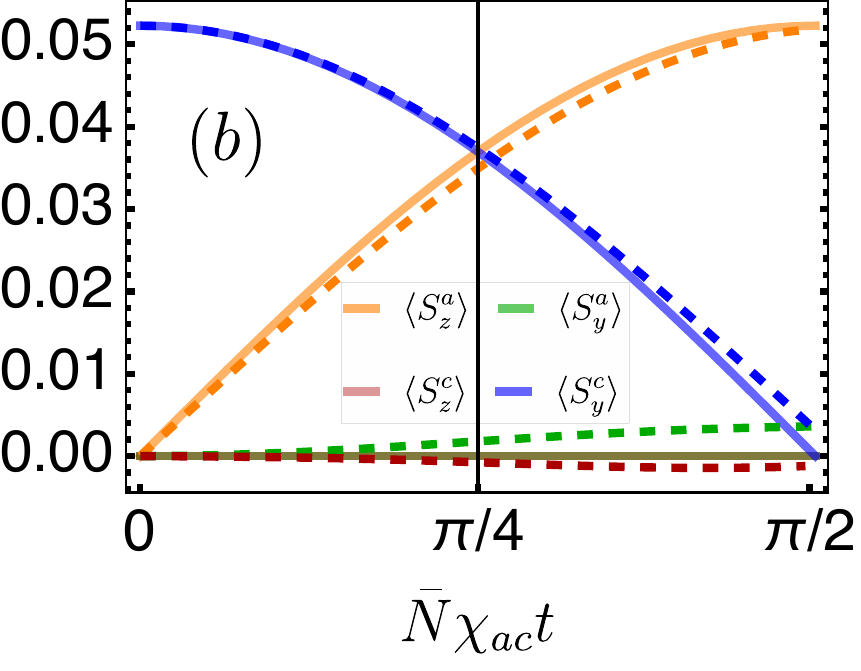}
	\caption{(a) Entanglement parameter $V_{s}$ (Eq.~\eqref{wit}) as a function of $r=\bar{N}\chi_{ab} t$ with $t$ the interaction time. Orange solid and green dashed lines represent the full model (FM) and effective spin model (ESM) dynamics. (b) Beam-splitter step: Dashed and corresponding solid lines show the dynamics obtained under the ESM  and HP, respectively. The vertical black lines in $(a)$ and $(b)$ correspond to interaction times where the TMS and BS operation are truncated, respectively.
Throughout our analysis, we assume parameters $\Omega_{a}=\Omega_{c}=-2\pi\times 19.1~ {\rm kHz},\Omega_{b}=-2\pi\times 18.8~ {\rm kHz}, ~  g_{a}=g_{b}=g_{c}=2\pi\times 3.6~ {\rm kHz}$, and $\delta_{M}/(2\pi)= -26~{\rm kHz}$. The dynamics is simulated with $N_{a}=N_{b}=N_{c}=\bar{N}=70$.}
	\label{fig:telepfig2}
\end{figure}

To mathematically formalize the excitation process,  we utilize Holstein–Primakoff (HP) transformation and approximate the collective spin operators by bosonic operators:
$\mathcal{\hat{S}}_{+}^{a}\simeq\sqrt{N_{a}} \hat{a}^{\dagger}$, $\mathcal{\hat{S}}_{+}^{c}\simeq\sqrt{N_{c}} \hat{c}^{\dagger}$ and $\mathcal{\hat{S}}_{+}^{b}\simeq\sqrt{N_{b}} \hat{b}$ up  to leading order in $1/N_{a,b,c}$,  such that $\mathcal{\hat{S}}_{x}^{l}\approx \sqrt{N_{l}/2}\hat{X}_{l}$, $\mathcal{\hat{S}}_{y}^{a,c}\approx-\sqrt{N_{a,c}/2}\hat{P}_{a,c}$ and  $\mathcal{\hat{S}}_{y}^{b}\approx\sqrt{N_b/2}\hat{P}_{b}$.  Here we have defined $\hat{X}_{l}=\frac{1}{\sqrt{2}}(\hat{l}+ \hat{l}^{\dagger}),\hat{P}_{l}=\frac{1}{i\sqrt{2}}(\hat{l}- \hat{l}^{\dagger}) $, which  satisfy  the standard commutation relation $[\hat{X}_{l},\hat{P}_{l}]=i$  for $l=a,b,c$ and  simplify   the spin-exchange Hamiltonian to a two-mode squeezing (TMS) interaction that generates correlations between two bosonic modes
\cite{SupplementalMaterials,PhysRevLett.130.113202} :
\begin{align}
\hat{H}_{\rm TMS}\approx -\chi_{ab}\bar {N} \pr{\hat{a}^{\dagger}\hat{b}^{\dagger}+\hat{b}\hat{a} }
\label{eq:HamoltoninaTMS}
\end{align} with $\bar {N} \sim N_a\sim N_b$.
The correlated creation of pair of spin excitations from the initial state results  in a thermofield double (TFD) state of the form $\textstyle{\ket{\psi_{ab}}=(1/\cosh{r})\sum_{n=0}^{\infty} (-i) ^n \tanh^n{r}\ket {n,n}}$. Here $r\equiv \bar{N}\chi_{ab} t$ is the magnitude of the two-mode squeezing parameter determined by the interaction time \cite{RevModPhys.84.621}.
The TFD state features an exponential growth and attenuation of the bosonic  hybrid quadratures defined as $\hat{X}^{\mp}\equiv (\hat{P}_{b}\pm\hat{X}_{a})/\sqrt{2}$ and $ \hat{P}^{\pm}\equiv(\hat{X}_{b}\pm\hat{P}_{a})/\sqrt{2} $, so 
$\hat{X}^{\mp}(t)= \hat{X}^{\mp}(0)e^{\pm \bar{N}\chi_{ab} t}$ and 
$\hat{P}^{\pm}(t)= \hat{P}^{\pm}(0)e^{\pm \bar{N}\chi_{ab} t}$. For $\chi_{ab}<0$ and in the limit $r\rightarrow -\infty$, i.e.  in the ideal case of infinitely large interaction time , one reaches  the EPR conditions \cite{Serafini2017,PhysRev.47.777}, ${\hat{P}}_{b}^{{\rm ideal}}=-{\hat{X}}_{a}^{{\rm ideal}}$ and ${\hat{X}}_{b}^{ {\rm ideal}}=-{\hat{P}}_{a}^{{\rm ideal}}$, or in terms of the spin variables (assuming the validity of HP with $\bar{N}\rightarrow \infty$ ), 
\begin{align}
\hat{S}_{y}^{{\rm ideal},b}-\hat{S}_{z}^{{\rm ideal},a}=0, \quad \hat{S}_{y}^{{\rm ideal},a}+\hat{S}_{z}^{{\rm ideal},b}=0
\label{eq:EPRProperty}
\end{align}
 Thus, in such an ideal limit, their variances $V[.]$ become negligible,  $V[\hat{S}_{y}^{{\rm ideal},b}-\hat{S}_{z}^{{\rm ideal},a}]\to 0$ and $ V[\hat{S}_{y}^{{\rm ideal},a}+\hat{S}_{z}^{{\rm ideal},b}]\to 0$.
Away from the ideal case,  the development of entanglement,  in terms of spin operators, can be witnessed by the inequality  
\begin{equation}
    V_s\equiv \frac{V[{\hat{S}}_{y}^{b}-{\hat{S}}_{z}^{a}]+V[{\hat{S}}_{z}^{b}+{\hat{S}}_{y}^{a}]}{ |\langle {\hat{S}}_{x}^{a}\rangle|+|\langle {\hat{S}}_{x}^{b}\rangle |}< 1 \label{wit}\end{equation} which serves as an entanglement witness \cite{PhysRevA.67.022320, PhysRevLett.84.2722,Julsgaard2001}. 
 In Fig.~\ref{fig:telepfig2} (a), we plot  $V_{s}$ with increasing $r$ both for the full model (FM) under the Hamiltonian in Eq.~\eqref{eq:FullDressedFrameHamlt} and the effective spin model (ESM) described by  Eq.~\eqref{eq:effenuclear_spin_spinHam1}. The entanglement between the $a$ and $b$ ensembles starts building as soon as the interaction becomes operational. The maximum entanglement is achieved at $r=r_{\rm min}\equiv \bar{N} \chi_{ab} t_{\rm TMS}$, the point  where $V_{s}$ is minimum. For $t>t_{\rm TMS} $ finite size effects start playing a role and as $V_{s}$ increases above  its minimum value, and it is no longer a useful quantifier of the entanglement. 

Following the BK teleportation scheme \cite{PhysRevLett.80.869}, setting    $t_{\rm TMS}$ when $V_s$ is optimal, the next step is to engineer an effective beam-splitter (BS) operation. We again start from the spin/phonon Hamiltonian in Eq.~\eqref{eq:FullDressedFrameHamlt} but 
 for this stage of the protocol, we set $\Omega_{b}=0$, since we want to freeze the dynamics in that state. Akin to the previous stage, we adiabatically eliminate the phonon mode and  go to a rotating frame, now set by $
U=e^{if_{r}\pc{\mathcal{\hat{S}}_{z}^{a}+\mathcal{\hat{S}}_{z}^{c}+\hat{m}^{\dagger}\hat{m}}t}$, where $f_r$, a frequency that depends on system parameters, can be chosen such that we obtain only a BS operation (see below). This rotating frame leads to an equation similar to Eq. \ref{eq:effenuclear_spin_spinHam1}, and we again obtain an effective spin-spin interaction, but with $c$ replacing $b$. In contrast to the prior  case, we now  want to initialize ensemble $c$ with a large spin projection  along the same direction as $a$, e.g. with most of the ions aligned along
 the south pole of the dressed $c$ Bloch sphere. Under this condition, when $\hat{H}_{ac}$  is applied to the  joint state, the flip-flop term 
$\mathcal{\hat{S}}_{-}^{a}\mathcal{\hat{S}}_{+}^{c}$ transfers a spin excitation
$\ket{\uparrow_a}$  in the $a$ ensemble to a spin $\ket{\uparrow_c}$ excitation in the $c$ ensemble and  vice-versa for the $\mathcal{\hat{S}}_{+}^{a}\mathcal{\hat{S}}_{-}^{c}$ term,  as desired for a BS. Note that in this case  the  self-interactions do not generate an energy penalty.  Nevertheless, they can induce a small self-generated precession of the collective spins in the $a,c$ ensembles that we want to ideally remove.  So we set $\Omega_a=\Omega_c$, and adjust $f_r$ to approximately cancel it
(see \cite{SupplementalMaterials}). In  the HP approximation limit,  the effective interaction between the $a$ and $c$ bosons reads now \cite{SupplementalMaterials}:
\begin{align}
\hat{H}_{\rm BS}=&-\chi_{ac}\bar{N}\pr{\hat{a}^{\dagger}\hat{c}+\hat{c}^{\dagger}\hat{a}},
\label{eq:BeamSpitterHam}
\end{align} Here  we defined the effective mode frequency $\Delta^{ac}_M =
\delta_M - f_r$ and assumed
$\chi_{aa} N_a\sim \chi_{cc} N_c\equiv \bar{\chi}\bar{N}$, and $N_a\sim N_c\sim~\bar{N}$ .
To realize  the required  50-50 BS operation,  the BS  Hamiltonian is applied for a time $\bar{N}\chi_{ac} t_{\rm BS}=\pi/4$  ( Fig.~\ref{fig:telepfig2} (b), i.e. when ensemble c is slightly displaced from the initial mean magnetization).

\begin{figure}[t!]
	\includegraphics[width=1\columnwidth]{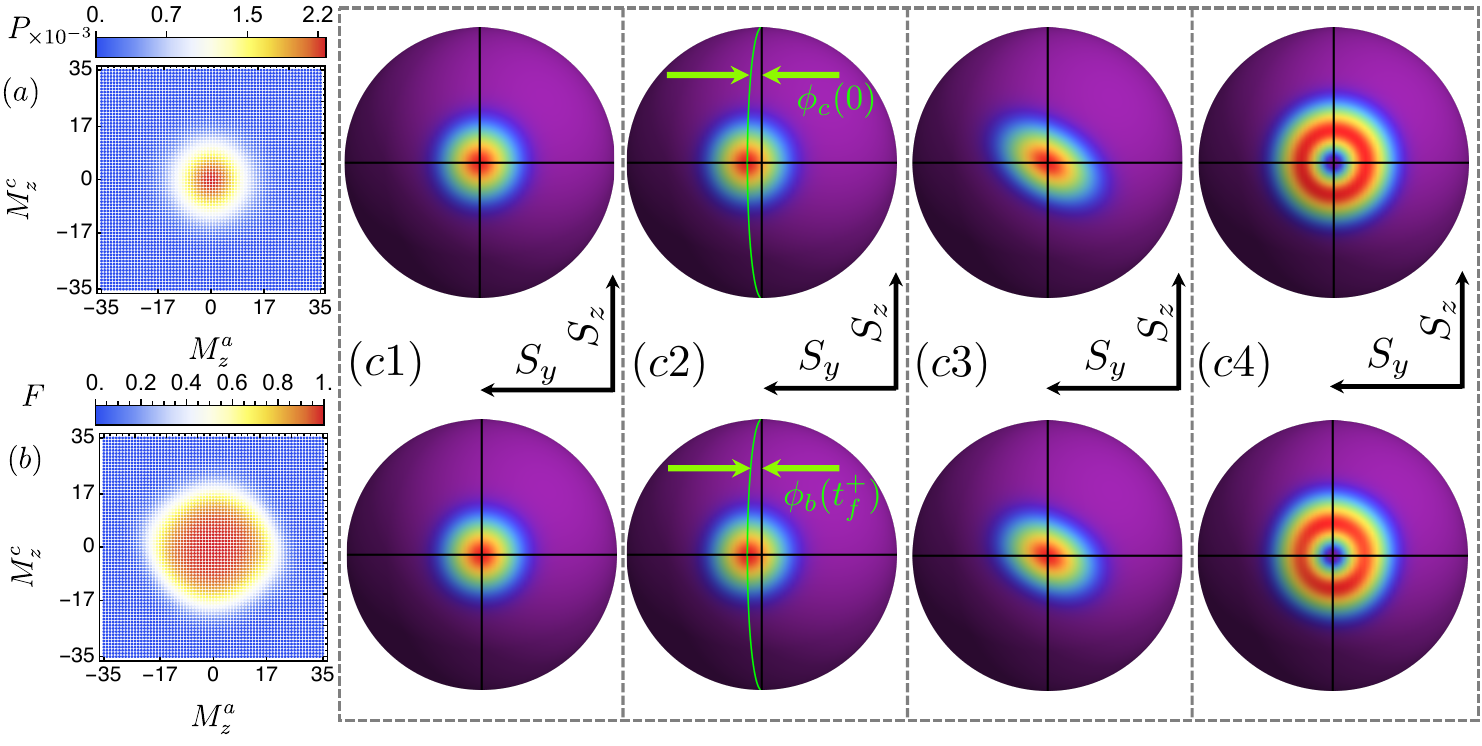}
	\caption{(a) Measurement outcome probability distribution function $P(M_z^{a},M_z^{c})$  of an input spin-coherent state. More examples are shown in \cite{SupplementalMaterials}. (b) Teleportation fidelity for different measurement outcomes. (c) Top row:  the Husimi-Q function of SC, PDSC, SS, and DS (with $k_{ c}=1$) input states, respectively. Bottom row: Husimi-Q function of teleported states. }
	\label{fig:telepfig3}
\end{figure}

As a result at  $t_f=t_{\rm BS}+t_{\rm TMS}$, we obtain  $ \hat{S}_{z}^{a}(t_f)=(\hat{S}_{y}^{c}(0)+\hat{S}_{z}^{a}({t_{\rm TMS}}))/\sqrt{2}$ and $\hat{S}_{z}^{c}(t_{f})=(\hat{S}_{z}^{c}(0)+ \hat{S}_{y}^{a}(t_{\rm TMS}))/\sqrt{2}$. Projective measurements to infer $z-$components of the $a$ and $c$ ensembles then cause a collapse according to $ \hat{S}_{z}^{a}(t_{\rm TMS})= \beta_{z} -\hat{S}_{y}^{c}(0)$ and  $ \hat{S}_{y}^{a}(t_{\rm TMS})= \beta_{y}  -\hat{S}_{z}^{c}(0)$, with  $\beta_{z}/\sqrt{2}=  M_{z}^{ a}\equiv (k_{a}-N_{a}/2) $ and  $\beta_{y}/\sqrt{2}=  M_{z}^{ c}\equiv (k_{c}-N_{c}/2)$, the measured outcomes with $k_{a,c}$ are the number of excitation in the spin ensembles. Ideally, due to the EPR property in Eq.~\eqref{eq:EPRProperty}, the state of ensemble $b$ is immediately projected according to, 
$\hat{S}_{y}^{{\rm ideal},b}(t_f^+)= \sqrt{2} M_{z}^{ a }(t_f) -\hat{S}_{y}^{c}(0) $ and 
$\hat{S}_{z}^{{\rm ideal},b}(t_f^+)=- \sqrt{2} M_{z}^{c } (t_f)+\hat{S}_{z}^{c} (0)$. 

These equations reflect that the projected state of ensemble $b$, is simply a ``rotated'' state of the input state of the ensemble $c$. By employing our knowledge of the measurement outcomes $M^{a,c}_z$, we can apply rotations of ensemble $b$ given by $\hat{U}_{b}=\hat{D}_{\pi}\hat{D}_{r}({\beta_{y}, \beta_{z}}) $, where $\hat{D}_{r}({\beta_{y}, \beta_{z}}) ={ \rm exp}[i(2/N_{b})(\beta_{z} \hat{S}_{z}^{b}+ \beta_{y} \hat{S}_{y}^{b}) ]$ \cite{SupplementalMaterials}, and $\hat{D}_{\pi}={\rm exp}(i\pi\hat{S}_{z}^{b})$, which  complete the desired teleportation. 

{\it Numerical Calculations: } We numerically simulate the many body spin-ensemble teleportation protocol  using exact diagonalization (ED). In Fig.~\ref{fig:telepfig3} (a) we show the results when the input state $\hat{\rho}_c(0)$ is a spin-coherent state for which the most probable outcome is  $\beta_{y}=0,~\beta_{z}=0$.
 To compare the teleported state $\hat{\rho}_{b}(t_{\it f}^+)$ to the input state for the most probable outcome, in Fig.~\ref{fig:telepfig3} (c), we plot the Husimi-Q functions, i.e.  $Q(\theta,\phi)\equiv(1/4{\pi})\langle\psi_{\rm SC}(\theta,\phi)|\hat{\rho}_c(0)|\psi_{\rm SC}(\theta,\phi)\rangle$  of  four different input states of $c$ given by: a spin coherent state 
$ \ket{\psi_{\rm SC}(\pi/2,\pi)} $ (SC), a phase-displaced spin coherent ${\rm exp}( -i \phi_c \hat{S}_{z}^{c} )\ket{\psi_{\rm SC}(\pi/2,\pi)}$(PDSC), a spin-squeezed state, $ \ket{\psi_{\rm SS}}={\rm exp}( -i \phi_{ss} (\hat{S}_{z}^{c})^2)\ket{\psi_{\rm SC}(\pi/2,\pi)}
$ (SS), and a  Dicke state, $\ket{\psi_{\rm \small{DS}}}={\rm exp}( -i [\pi/2] \hat{S}_{y}^{c} )\hat{S}_{+}^c\ket{\psi_{\rm SC}(\pi,0)}$(DS) with one excitation. Here $\ket{\psi_{\rm SC}(\theta,\phi)}$ represents a generic spin coherent state \cite{Ma2011}. The Dicke state  is included here to show that the teleportation protocol applies also for such states, while we note that their preparation as input states would need a higher order non-linearity or a heralding protocol \cite{McConnell2015}. We also show (bottom row) their teleported versions. For all cases, the mean orientations and noise distributions of the teleported states match the ones of the input. To make a more quantitative comparison, we compute the Uhlmann fidelity $\textstyle{F (M_z^a,M_z^c)=\pr{{\rm Tr}\pc{\sqrt{\sqrt{\hat{\rho}_{c}(0)}\hat{\rho}_{b}(t_{\it f}^+)\sqrt{\hat{\rho}_{c}(0)}}}}^2 }$ \cite{Jozsa1994} between the input and teleported SC states  for different measurement outcomes and show it in Fig.~\ref{fig:telepfig3} (b) for $N_{a}=N_{b}=N_{c}=\bar{N}=70$.  Similar results are shown for PDSC, SS, and DS states  in \cite{SupplementalMaterials}.

 For the most probable outcomes $\tilde M_z^a,\tilde M_z^c$, the   fidelities are given by $F_{\rm SC}(\tilde M_z^a,\tilde M_z^c)=F_{\rm PDSC}(\tilde M_z^a,\tilde M_z^c)=0.99$,  $F_{\rm SS}(\tilde M_z^a,\tilde M_z^c)=0.98$, and $F_{\rm DS}(\tilde M_z^a,\tilde M^z_c)=0.99$. With the current protocol, for an input phase $\phi_{c}(0)=6 \degree$, we obtain a teleported state with a  phase $\phi_{b}(t_{f}^+)=5.34 \degree$. For the SS states we obtain a squeezing parameter $\xi_{b}(t_{ f}^+)=-3.17({\rm dB})$, for an  input state with $\xi_{c}(0)=-4.15 ({\rm dB})$. The errors are limited by curvature corrections due to finite ion number.
 For the average fidelity $\textstyle{\bar F=\sum_{M_z^a,M_z^c}P(M_z^a,M_z^c)F({M_z^a,M_z^c)}}$ we obtain the averages fidelities to be $\bar F_{\rm SC}=\bar F_{\rm PDSC}=~0.87$, $\bar F_{\rm SS}=0.85$, and $\bar F_{\rm DS}=0.68$ for the corresponding input states with $\bar{N}=70$.

\begin{figure}[t]
	\includegraphics[width=0.46\columnwidth]{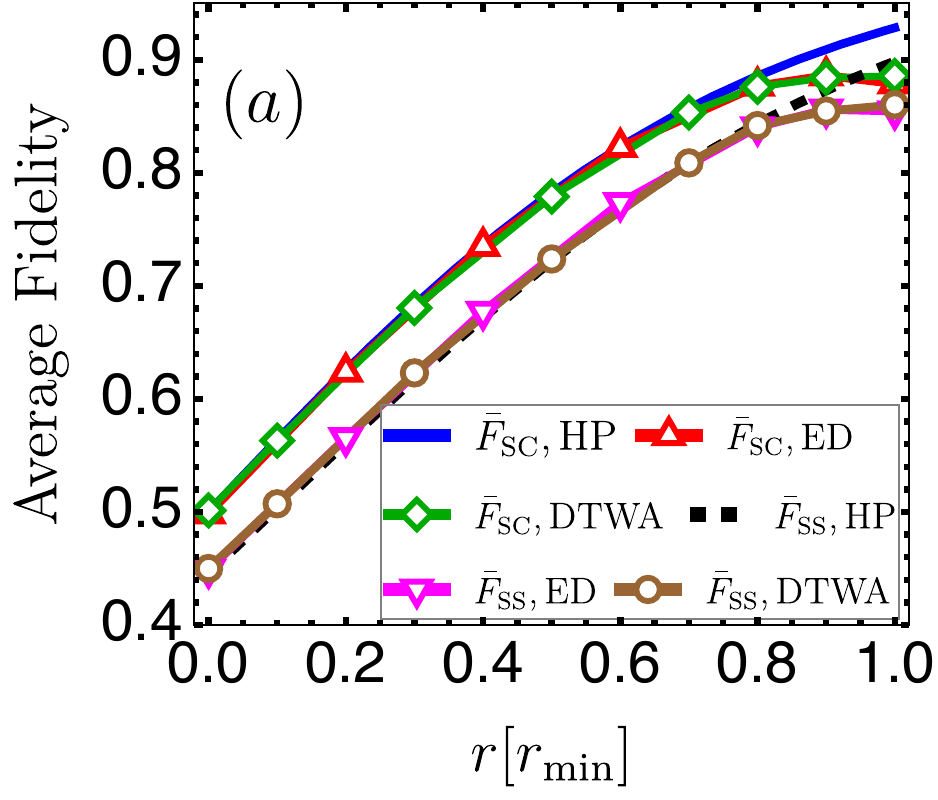}
	\includegraphics[width=0.48\columnwidth]{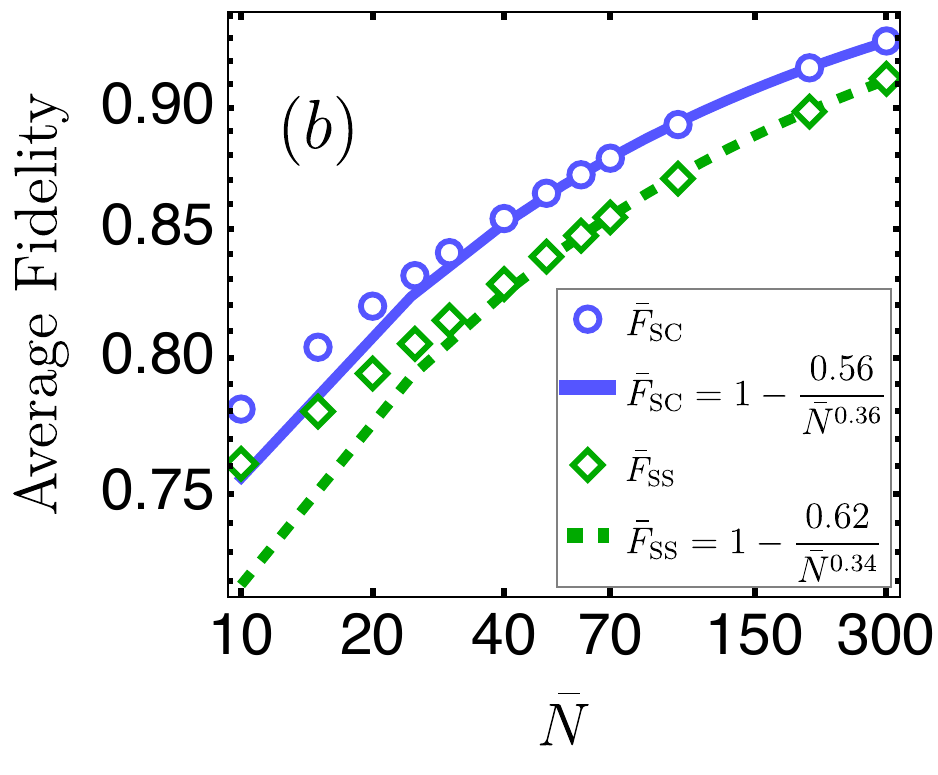}
	\caption{(a) Numerically computed average teleportation fidelity for SC and SS input states as a function of entangling interaction time up to the time when entanglement is maximum with $\bar{N}=70$.  (b) The average teleportation fidelity scaling with the number of ions $\bar{N}$ for SC and SS input states.}
	\label{fig:telepfig4}
\end{figure}

 To further assess how the available entanglement affects the performance of the teleportation, we plot the average fidelity as a function of $r$ up to $r_{\rm min}$ for a fixed number of ions $N_{l}$ as shown in Fig.~\ref{fig:telepfig4} (a). We numerically average the fidelity both with the ED and the discrete truncated Wigner approximation (DTWA) methods \cite{SupplementalMaterials}. As expected, there is a monotonic increase in the average fidelity against $r$ both for the SC and SS states and the fidelity closely follows the analytic expression obtained with the HP solution \cite{PhysRevA.78.062340,Li2015/04,PhysRevA.65.022310,SupplementalMaterials} at short times. In Fig.~\ref{fig:telepfig4} (b) we compute the average fidelity scaling with $\bar{N}$ for the SC and SS cases. We find a scaling  $\bar{F}_{\rm SC} \simeq 1-[0.56/(\bar{N}^{0.36}) ]$ and $\bar{F}_{\rm SS} \simeq 1-[0.62/(\bar{N}^{0.34}) ]$. 

\textit{Experimental considerations:} Many of the necessary  ingredients  for the protocol have been individually demonstrated already in Penning traps:  Population of various nuclear spins levels using rf-pulses have  been  achieved  in $^9$Be$^+$\cite{PhysRevA.84.012510}. Global sub-ensemble  rotations enjoy fidelities as high as  $99\%$.  The COM mode has been  cooled down  close to its ground state  value \cite{PhysRevLett.122.053603} and  the observed  phonon damping rate $\gamma_{m}$ is mostly irrelevant. Since the  dominant decoherence channel  is single spin  dephasing induced by Rayleigh and Raman scattering with rates  $\Gamma_{\rm Rl}>\Gamma_{\rm Rm}$ \cite{PhysRevLett.121.040503} the main limitation is the  requirement   $\Gamma_{\rm Rl} t_{\rm TMS} < V_s$,  $t_{\rm BS} \Gamma_{\rm Rl}\ll 1 $  and
$t_{f} \Gamma_{\rm Rl}<\xi_{c}(0)$. For current experimental parameters  $g_a\sim g_b\sim g_c \approx 2\pi \times 4~{\rm k Hz}$, $\bar{N}\approx70$ and $\Gamma_{\rm Rl}\approx 250 {\rm\text{ } s}^{-1}$, we expect to  achieve a TMS parameter $r\approx0.55$ and therefore an average teleportation fidelity of $\bar{F}\approx 0.75$ for a coherent state and $\bar{F}\approx 0.65$ for a SS state with $\xi_{c}(0)\approx -5.3~\text{dB}$. 

Fluorescence measurements have been also demonstrated  \cite{Bohnet2016} for a single nuclear spin component $m_{I}=\pm3/2$  prepared via optical pumping \cite{Britton2012,Wineland1980}. Even though the collection efficiency in current experiments may be just at the threshold needed to avoid errors from off resonant light  scattering into other nuclear spin states during detection, in the future,  improved detection efficiency can be gain by using high-aperture lenses or a build-up cavity. Alternatively, instead
of nuclear spins, an additional  narrow
optical transition enabled by optical-metastable-ground-states (OMG) \cite{Allcock2021,Yang2022,Bzvan2023,McMahon}, could significantly enhance the detection fidelity.
For example, instead of  $^9{\rm Be}^{+}$, one can use $^{43}{\rm Ca^{+}}$, and during the measurement step, transfer the qubits to the $^{2}{\rm D}_{5/2} $ level.

\textit{Outlook:} While we have  focused on using only  internal levels as effective bosonic modes, the long coherence times  and cooling capability of the phonon modes in current experiments also offers the possibility of taking advantage of them as extra active channels when  operating close to resonance. The advantage is  faster preparation time scales and thus less sensitivity to decoherence with the only overhead  being   an  additional transfer operation between the phonon mode and  a spin degree of freedom for readout \cite{lewisswan2023exploiting}. 
To perform teleportation between spatially separated ensembles, we could take advantage of crystal bilayers or 3D crystals 
\cite{hawaldar2024bilayer} and use   addressable sections of the crystal as collective spin degrees of freedom.
In the future, the opportunity of performing  teleportation and scrambling operations \cite{Grttner2017} in these system and the existence of at least five effective channels, four nuclear spin sub-ensembles and one active phonon mode, opens also the exciting possibility to implement protocols to disentangle scrambling and decoherence via quantum teleportation \cite{Landsman2019,Blok2021,Yoshida2019}. Beyond teleportation, 
similar schemes could be used for retroactive squeezing generation  for enhanced displacement sensing in the Penning trap \cite{Bao2020}.\\
\textit{Acknowledgments:}  We thank  Raphael Kaubruegger and Jenny Lilieholm for their useful feedback on the manuscript and David Wellnitz  for  helpful discussions.  This material is based upon work supported by the  Heising-Simons  foundation  and the U.S. Department of Energy, Office of Science, National Quantum Information Science Research Centers, Quantum Systems Accelerator. We also acknowledge funding support from  ARO grant  W911NF24-1-0128, the NSF JILA-PFC PHY-2317149  and  NIST. KM acknowledges the Carlsberg Foundation through the ``Semper Ardens'' Research Project QCooL.

$^*$Now at Rigetti Computing, 775 Heinz Avenue, Berkeley, California 94710, USA.
\newpage
\bibliographystyle{apsrev4-1}
\bibliography{References.bib}
\widetext
\newpage
\begin{center}
\textbf{\large Supplemental Materials: Generating Einstein–Podolsky–Rosen correlations for  teleporting collective spin states in a two dimensional trapped ion crystal}
\end{center}
\setcounter{equation}{0}
\setcounter{figure}{0}
\setcounter{table}{0}
\setcounter{page}{1}
\makeatletter
\renewcommand{\theequation}{S\arabic{equation}}
\renewcommand{\thefigure}{S\arabic{figure}}
\renewcommand{\bibnumfmt}[1]{[S#1]}
\renewcommand{\citenumfont}[1]{S#1}
This supplemental material is organized as follows: In Sec.~\ref{Sup:Section1}, we describe the method for state initialization of three spin ensembles. In Sec.~\ref{Sup:Section2}, we derive the effective spin exchange interaction from the full model following the constraints on the system parameters. We further derive the effective form of the spin-spin interaction that allows us to identify the spin EPR variables. Additionally, we  derive the corresponding  two-mode squeezing and beam-splitter interaction in the  Holstein–Primakoff (HP) picture. In Sec.~\ref{Sup:Section3}, we describe the dynamics in the HP picture to study the corresponding average teleportation fidelity. In Sec.~\ref{Sup:Section4}, we report the mean-field equations of motion for the full system. These are used to simulate the full model (FM) dynamics using the truncated approximation (TWA) method.  In Sec.~\ref{Sup:Section5}, we derive the effective displacement operator that is applied to ensemble $b$ locally to complete the teleportation protocol. Finally in Sec.~\ref{Sup:Section6}, we include additional  supportive plots. These include  measurement outcome probabilities  and fidelity distribution functions for all the input states to be teleported, as discussed in the main text. We also include the magnetization statistics of the input and the teleported states, when the state to be teleported is an  entangled spin-squeezed state. We additionally demonstrate the teleportation of Dicke states with two spin excitation, $k_c=2$. Moreover, we show the average teleportation fidelity against the number of ions, when the input states are Dicke state with one and two spin excitation, i.e. with $k_c=1, ~k_c=2$.
 \section{State Initialization }\label{Sup:Section1}
%
Here we discuss the initial state preparation of three ensembles as shown in Fig.~\ref{fig:telepfig1} of the main text. The preparation stage of the three ensembles relies on initialization $\bar{N}=(1/3) N$, i.e. one-third of the ions to be in each spin ensembles $a,b, c$. This can be established by initializing the ions in nuclear spin $l=a$, followed by applying  RF-pulse(s) to  transfer equal averaged population in the state $\ket{e_{l}}$, for $l={a,b,c}$. Any coherence developed between the nuclear spin during the transfer stage can be suppressed via optically pumping  since they are fully decoupled in the absence of RF pulses. Once one-third of the ions are prepared in each of the nuclear spin excited states $\ket{e_{l}}$, microwave  rotations $\hat{U}_{l}=e^{i(\pi/2)\hat{S}_{y}^l}$ for $l=a,c$  can be used to  prepare these ensembles along the $-x$ direction, while the use of a microwave with opposite phase $\hat{U}_{b}=e^{-i(\pi/2)\hat{S}_{y}^b}$ would allow preparation of ensemble $b$ along the $+x$ direction.
\section{Derivation of effective bosonic mode operations between different subsystems}\label{Sup:Section2}
In this section we analyze the dynamics in the dressed frame and derive the effective spin-spin interactions between different sub-systems for the various stages of the protocol as stated in the main text. Moreover, we derive how these spin-spin interactions lead to the desired  two-mode squeezing and beam-splitting operations. 
\subsection{Entangling operation  between $a$ and $b$ ensembles}\label{Appendix_E}
In the dressed basis, the Hamiltonian of the total system is given by Eq.~\eqref{eq:FullDressedFrameHamlt} of the main text. This Hamiltonian, when the terms $\sim \epsilon_{l} (\hat{m}+\hat{m}^\dagger)$ are included due to the small imbalance in polarizations of the Raman beams, is given by
\begin{align}\label{eqApp:FullDressedFrameHamlt}
\hat{H}=&\sum_{l=a,b}\Omega_{l}\mathcal{\hat{S}}_{z}^{l}-\sum_{l=a,b,c}\pr{\mathcal{G}_{l}(\hat{m}+\hat{m}^{\dagger})(\mathcal{\hat{S}}_{x}^{l}+\epsilon_{l}\hat{I}_l)}+\delta_{M}\hat{m}^{\dagger}\hat{m}, \quad {\rm where}\quad  \mathcal{G}_{l}=g_{l}/\sqrt{N}.
\end{align}

\subsubsection{Tavis-Cumming interaction between phonon mode and ensembles a and b}
To perform the first stage of the protocol of the teleportation we set $\Omega_{c}=0$. Furthermore, we consider particular values of the microwave derive $\Omega_{a}=\Omega_{\rm av}+\delta_{ab}$  and $\Omega_{b}=\Omega_{\rm av}-\delta_{ab}$, where $\Omega_{\rm av}=\pc{\Omega_{a}+\Omega_{b}}/2$ and  $\delta_{ab}=\pc{\Omega_{a}-\Omega_{b}}/2$ \cite{PhysRevLett.130.113202}. These parameters are merely controlled by the local microwaves' Rabi frequencies. For this choice of dressed states energy splittings, we move into the interaction picture both for spins and the phonon with respect to average value $\Omega_{\rm av}$. This is  performed by the transformation
\begin{align}
U=e^{-i\Omega_{\rm av}\pc{\mathcal{\hat{S}}_{z}^{a}+\mathcal{\hat{S}}_{z}^{b}+\hat{m}^{\dagger}\hat{m}}t}.
\end{align}
In this interaction picture, the Hamiltonian (valid for $\{\delta_{ab},\mathcal{G}_{a}\sqrt{N_{a}},\mathcal{G}_{b}\sqrt{N_{b}},\mathcal{G}_{c}\sqrt{N_{c}},\Delta_{M}^{ab} \}\ll \Omega_{\rm av}$ ) is given by
\begin{align}\label{EqApp:HamolDressed}
\hat{H}=\Delta_{M}^{ab}\hat{m}^{\dagger}\hat{m}-\sum_{l= a,b}\frac{\mathcal{G}_{l}}{2}\pc{\mathcal{\hat{S}}_{+}^{l}\hat{m}+\hat{m}^{\dagger}\mathcal{\hat{S}}_{-}^{l}}+\delta_{ab}\pc{\mathcal{\hat{S}}_{z}^{a}-\mathcal{\hat{S}}_{z}^{b}},
\end{align}
where $\Delta_{M}^{ab}\equiv\delta_{M}-\Omega_{\rm av}$. Note that the spin-phonon interaction for ensemble $c$ is rotated out in this interaction picture due to fact that phonon annihilation and creation operators accumulate the fast oscillating phase factor as determined by $e^{-i \Omega_{ \rm av}t}$ and  $e^{i\Omega_{\rm av}t}$ respectively. In addition, the phonon annihilation and creation operators in the local displacement caused by the imbalance in polarizations of the Raman beams also accumulate the same fast oscillating phase factors. Therefore the terms proportional to $\epsilon_{l}$ are rotated out.  

\subsubsection{Adiabatic elimination of the phonon mode to obtain effective spin-spin interaction for ensembles a and b}
We now assume that the phonon mode is far-detuned and performs fast oscillation on the time scale of $(\Delta_{M}^{ab})^{-1}$ compared to the time scale of the internal spin dynamics. The separation of time scales between these subsystem is fulfilled by the conditions $\{g_{a}/\sqrt{N_a} ,  g_{b}/\sqrt{N_b}\}<4\Delta_{M}^{ab}$. In this case, the phonon dynamics can be averaged out to zero on the time scale of the spin dynamics and we can set  $\partial_{t}\hat{m}=0$. From Eq.~\eqref{EqApp:HamolDressed}, the Heisenberg equation of motion for the phonon annihilation operator is
\begin{align}
\partial_{t}\hat{m}=-i\Delta_{M}^{ab}\hat{m}+i\sum_{l=a,b}\frac{\mathcal{G}_{l}}{2}\hat{S}_{-}^{l}
\end{align}
By setting $\partial_{t}\hat{m}=0$ we get
\begin{align}
\hat{m}\simeq \sum_{l=a,b}\frac{\mathcal{G}_{l}}{2\Delta_{M}^{ab}}\hat{S}_{-}^{l}, \quad \hat{m}^{\dagger}\simeq \sum_{l=a,b}\frac{\mathcal{G}_{l}}{2\Delta_{M}^{ab}}\hat{S}_{+}^{l}.
\end{align}
Thus the phonon mode adiabatically follows the spins. Inserting back these expression in  Eq.~\eqref{EqApp:HamolDressed}, we get an effective spin Hamiltonian  given by
\begin{align}\label{eqApp:EffectiveSpinModel}
\hat{H}_{ab}=-\frac{1}{4\Delta_{M}^{ab}} \pc{\mathcal{G}_{a} \mathcal{\hat{S}}_{+}^{a}+ \mathcal{G}_{b}\mathcal{\hat{S}}_{+}^{b}} \pc{\mathcal{G}_{a} \mathcal{\hat{S}}_{-}^{a}+ \mathcal{G}_{b}\mathcal{\hat{S}}_{-}^{b}}
+\delta_{ab}\pc{\mathcal{\hat{S}}_{z}^{a}-\mathcal{\hat{S}}_{z}^{b}}.
\end{align}
This is the effective spin model that we state in Eq.~\eqref{eq:effenuclear_spin_spinHam1} of the main text. In the main text we numerically show that, strictly following the conditions on the system parameters as stated above, the dynamics of the full model (FM) as given by Eq.~\eqref{eqApp:FullDressedFrameHamlt} and of the effective spin model (SM) given by Eq.~\eqref{eqApp:EffectiveSpinModel}, showcase equivalent dynamics. Therefore in this first stage of the protocol, we have a pure entangling interaction between ensembles $a$ and $b$, while the phonon mode and third spin ensemble $c$ do
 not participate in the dynamics.
\subsubsection{Schwinger boson representation and two-mode squeezing (TMS) operation between $a$  and $b$ ensembles}
In the above equation, we write the collective spin operator in terms of Schwinger bosons as:
\begin{align}
&\mathcal{S}_{+}^{a}=\hat{a}^{\dagger}_{\uparrow}\hat{a}_{\downarrow}, \quad  \mathcal{S}_{+}^{b}=\hat{b}^{\dagger}_{\uparrow}\hat{b}_{\downarrow},\\
&\mathcal{S}_{z}^{a}=\frac{1}{2}\pc{\hat{a}^{\dagger}_{\uparrow}\hat{a}_{\uparrow}-\hat{a}^{\dagger}_{\downarrow}\hat{a}_{\downarrow}}, \mathcal{S}_{z}^{b}=\frac{1}{2}\pc{\hat{b}^{\dagger}_{\uparrow}\hat{b}_{\uparrow}-\hat{b}^{\dagger}_{\downarrow}\hat{b}_{\downarrow}}.
\end{align}
The resulting interaction between the boson modes includes four wave mixing processes and is highly nonlinear. However, the ensembles $a$ and $b$ are initially spin-polarized along the  $-x$ and $+x$ directions, respectively, and the states associated with the bosonic operators $\hat{a}_{\downarrow} $, $\hat{a}^{\dagger}_{\downarrow} $,  $\hat{b}_{\uparrow} $, $\hat{b}^{\dagger}_{\uparrow} $ are macroscopically populated. For short interaction times, we can to a good approximation, replace these operators with their mean values as  $\hat{a}_{\downarrow}=\sqrt{N_{a}},\hat{b}_{\uparrow}=\sqrt{N_{b}} $ (i.e. Holstein–Primakoff approximation for $N_{l}\gg1$). This allows us to rewrite the Eq.~\eqref{eqApp:EffectiveSpinModel} as:
\begin{align}
\hat{H}_{ab}=-\frac{1}{4\Delta_{M}^{ab}}\pr{N_{a}\mathcal{G}_{a}^{2}\hat{a}^{\dagger}_{\uparrow}\hat{a}_{\uparrow}+N_{b}\mathcal{G}_{b}^{2}\hat{b}^{\dagger}_{\downarrow}\hat{b}_{\downarrow}}+\delta_{ab}\pr{\hat{a}^{\dagger}_{\uparrow}\hat{a}_{\uparrow}+\hat{b}^{\dagger}_{\downarrow}\hat{b}_{\downarrow}}-\frac{\mathcal{G}_{a}\mathcal{G}_{b}\sqrt{N_{a}N_{b}}}{4\Delta_{M}^{ab}}\pr{\hat{a}^{\dagger}_{\uparrow}\hat{b}^{\dagger}_{\downarrow}+\hat{b}_{\downarrow}\hat{a}_{\uparrow}}
\label{eq:SuppTMS}
\end{align}
The first two terms of the Hamiltonian represent the self-interactions, that translate  into  energy shifts. The third term in the Hamiltonian represents the correlated creation (transfer) of particles in states $\ket{\uparrow_{a}}$ and $\ket{\downarrow_{b}}$, i.e. along the $+x$ and $-x$ direction respectively, as represented on the Bloch spheres shown in the main text. We define $\chi_{\alpha \alpha^{\prime}}\equiv (\mathcal{G}_{\alpha}\mathcal{G}_{\alpha^{\prime}})/4\Delta_{M}^{ab}$, with $\{\alpha ,\alpha^{\prime} \} \in \{a,b\}$. In order to cancel the self-interaction terms, we set   $\chi_{aa}N_{a}=\chi_{bb}N_{b}=\chi\bar{N}=\delta_{ab}$. Moreover omitting the redundant labeling $\{\uparrow,\downarrow\}$, we get the final form of the Hamiltonian given by
\begin{align}
\hat{H}_{\rm TMS}=&-\chi_{ab}\bar{N} \pr{\hat{a}^{\dagger}\hat{b}^{\dagger}+\hat{b}\hat{a} }.
\label{eq:nuclear_spin_BosonicformTMS}
\end{align}
This is the effective two mode squeezing Hamiltonian stated in the main text.

\subsubsection{Derivation for effective spin-spin  entangling interaction between $a$ and $b$ ensembles }\label{Appendix_C}

In this section, we derive the  form of the effective spin-spin entangling interaction that allows us to obtain the EPR variables in terms of the collective spin operators. We rewrite the Hamiltonian in Eq.~\eqref{eqApp:EffectiveSpinModel} as
\begin{align}
\hat{H}_{ab}=-\frac{1}{4\Delta_{M}^{ab}} \pr{\mathcal{G}_{a}^2(\mathcal{\hat{S}}^{a^2}-{\mathcal{\hat{S}}_{z}^{a^2}}+\mathcal{\hat{S}}_{z}^a)+\mathcal{G}_{b}^2(\mathcal{\hat{S}}^{b^2}-{\mathcal{\hat{S}}_{z}^{b^2}}+\mathcal{\hat{S}}_{z}^b)}-\frac{1}{4\Delta_{M}^{ab}}\pr{2\mathcal{G}_{a}\mathcal{G}_{b}(\mathcal{\hat{S}}_{x}^{a}\mathcal{\hat{S}}_{x}^{b}+\mathcal{\hat{S}}_{y}^{a}\mathcal{\hat{S}}_{y}^{b}) }+\delta_{ab}\pc{\mathcal{\hat{S}}_{z}^{a}-\mathcal{\hat{S}}_{z}^{b}}
\label{eqAPP:effenuclear_spin_spinHamDresedframe}
\end{align}
We now perform linearization of the spin operator $\mathcal{\hat{S}}_{z}^{l^2} ~(l=\{a,b\})$ by writting their fluctuations on top of initial mean field values as $\mathcal{\hat{S}}_{z}^{l}=\langle \mathcal{\hat{S}}_{z}^{l}\rangle+\delta \mathcal{\hat{S}}_{z}^{l}$. According to initial conditions, $\langle \mathcal{\hat{S}}_{z}^{a}\rangle\sim -N_a/2$ and  $\langle \mathcal{\hat{S}}_{z}^{b}\rangle\sim N_b/2$, and $|\langle\mathcal{\hat{S}}_{z}^{a,b}\rangle|\gg 1$, we ignore the non-linear interaction terms $\sim \delta\mathcal{\hat{S}}_{z}^{l} \delta\mathcal{\hat{S}}_{z}^{n}$ and terms $\mathcal{O}(1) $. We then  obtain the Hamiltonian
\begin{align}\label{eqAPP:nuclear_spin_intraction2}
\hat{H}_{ab}=-\frac{N_a \mathcal{G}_{a}^2}{4\Delta_{M}^{ab}}\mathcal{\hat{S}}_{z}^{a}+\frac{N_b \mathcal{G}_{b}^2}{4\Delta_{M}^{ab}}\mathcal{\hat{S}}_{z}^{b}+\delta_{ab}\pc{\mathcal{\hat{S}}_{z}^{a}-\mathcal{\hat{S}}_{z}^{b}}
-2\chi_{ab}\pr{\mathcal{\hat{S}}_{x}^{a}\mathcal{\hat{S}}_{x}^{b}+\mathcal{\hat{S}}_{y}^{a}\mathcal{\hat{S}}_{y}^{b}},
\end{align}
where we disregard the constant energy terms. Here $\chi_{ab}=(\mathcal{G}_a \mathcal{G}_b)/(4\Delta_{M}^{ab})$. For the resonant case $(N_a \mathcal{G}_{a}^2)/(4\Delta_{M}^{ab})=(N_{b}\mathcal{G}_{b}^2)/(4\Delta_{M}^{ab})=\delta_{ab}$, where self-interaction terms are canceled, we get the Hamiltonian 
\begin{align}\label{eq:nuclear_spin_intraction3}
\hat{H}=
-2\chi_{ab}\pr{\mathcal{\hat{S}}_{x}^{a}\mathcal{\hat{S}}_{x}^{b}+\mathcal{\hat{S}}_{y}^{a}\mathcal{\hat{S}}_{y}^{b}},
\end{align}
which we write in the factorized form as:
\begin{align}\label{eqAPP:spinsqueezingeffetiveEq4}
\hat{H}_{ab}=-\chi_{ab}(\mathcal{\hat{S}}_{y}^{b}+\mathcal{\hat{S}}_{x}^{a})(\mathcal{\hat{S}}_{x}^{b}+\mathcal{\hat{S}}_{y}^{a})+\chi_{ab}(\mathcal{\hat{S}}_{y}^{b}-\mathcal{\hat{S}}_{x}^{a})(\mathcal{\hat{S}}_{x}^{b}-\mathcal{\hat{S}}_{y}^{a}).
\end{align}
The above equation can also be written in the lab-frame as
\begin{align}\label{eq:nuclear_spin_intraction4t}
\hat{H}_{ab}=-2\chi_{ab}\pr{\hat{S}_{z}^{a}\hat{S}_{z}^{b}+\hat{S}_{y}^{a}\hat{S}_{y}^{b}},
\end{align}
which is the effective collective entangling interaction as shown in the Fig.~\ref{fig:telepfig1} (c) of the main text. Note that the HP representation allows us to map the collective spin component on the bosonic according to $\textstyle{\hat{\mathcal{S}}_{x}^{a}=-\hat{S}_{z}^{a}= \sqrt{(N_a/2)}\hat{X}_{a}}$, $\textstyle{\hat{\mathcal{S}}_{y}^{a}=\hat{S}_{y}^{l}=-\sqrt{(N_a/2)}\hat{P}_{a}}$. And for ensemble $b$, we have $\textstyle{\hat{\mathcal{S}}_{x}^{b}=-\hat{S}_{z}^{b}=\sqrt{(N_b/2)}\hat{X}_{b}}$ and  $\textstyle{\hat{\mathcal{S}}_{y}^{b}=\hat{S}_{y}^{b}=\sqrt{(N_b/2)}\hat{P}_{b}}$.
Therefore, by inserting these relations in Eq.\eqref{eq:nuclear_spin_intraction4t} and considering $N_{a}=N_{b}=\bar{N}$, one gets the two mode squeezing interaction as stated in Eq. \eqref{eq:nuclear_spin_BosonicformTMS}.

Eq. \eqref{eqAPP:spinsqueezingeffetiveEq4} reflects the spin squeezing for the set of variables $\{ \mathcal{\hat{S}}_{y}^{b}+\mathcal{\hat{S}}_{x}^{a},\mathcal{\hat{S}}_{x}^{b}-\mathcal{\hat{S}}_{y}^{a}\}$ for negative $\chi_{ab}$, or $\{\mathcal{\hat{S}}_{x}^{b}+\mathcal{\hat{S}}_{y}^{a}, \mathcal{\hat{S}}_{y}^{b}-\mathcal{\hat{S}}_{x}^{a}\}$  for positive $\chi_{ab}$ \cite{PhysRevLett.130.113202}. In the lab frame, the corresponding EPR variables are $\{S_{y}^{b}-S_{z}^{a}, S_{z}^{b}+S_{y}^{a}\} $ or $\{S_{y}^{a}-S_{z}^{b}, S_{y}^{b}+S_{z}^{a}, \}$ for negative $\chi_{ab}$ or positive $\chi_{ab}$, respectively. In the main text we choose EPR variable with negative $\chi_{ab}$ given their convenience to implement teleportation circuit. By choosing so, the measurement process maps to measuring the $M_{z}^a$ and $M_{z}^c$ magnetization without performing extra rotations before the measurement process.
\subsection{Beam-Splitter operation between $a$ and $c$ ensembles}\label{Appendix_F}
In order to establish  the second stage of the teleportation protocol, we perform a beam-splitter type interaction between the ensembles $a$ and $c$. In order to to establish this, we once again start from lab frame Hamiltonian Eq.~\eqref{eq:FullDressedFrameHamlt} from the main text, where we also include the terms proportional to $\epsilon_{l}$ that cause local displacements of the mode that commonly interact with each of the spin ensembles. We consider that during this stage of the protocol, we turn off the drive to the ensemble $b$ by setting $\Omega_{b}=0$. Nevertheless, we propose to drive the ensemble $a$  and $c$  by the same microwave drive strength i.e. $\Omega_{a}=\Omega_{c}\equiv\Omega$. 
\subsubsection{Tavis-Cumming interaction between phonon mode, and $a, c$ ensembles }
We follow the same procedure as before and
move into an  interaction picture both for spins and phonon in this case using the transformation
\begin{align}
U=e^{-if_{r}\pc{\mathcal{\hat{S}}_{z}^{a}+\mathcal{\hat{S}}_{z}^{c}+\hat{m}^{\dagger}\hat{m}}t},
\end{align} with $f_{r}$ found in a self consistent manner as shown below, under which  the Hamiltonian is given by
\begin{align}\label{Eq:HamolDressed}
\hat{H}_{ac}=\Delta_{M}^{ac}\hat{m}^{\dagger}\hat{m}-\sum_{a,c}\mathop{}_{\mkern-5mu l}\frac{\mathcal{G}_{l}}{2}\pc{\mathcal{\hat{S}}_{+}^{l}\hat{m}+\hat{m}^{\dagger}\mathcal{\hat{S}}_{-}^{l}}+\delta_{ac}\pc{\mathcal{\hat{S}}_{z}^{a}+\mathcal{\hat{S}}_{z}^{c}},
\end{align}
where $\Delta_{M}^{ac}\equiv\delta_{M}-f_{r}$ and $\delta_{ac}=\Omega-f_{r}$. The above Hamiltonian is valid provided that $\{\delta_{ac},\mathcal{G}_{a}\sqrt{N_{a}},\mathcal{G}_{b}\sqrt{N_{b}},\mathcal{G}_{c}\sqrt{N_{c}},\Delta_{M}^{ac}\}\ll f_{r}$. Here, akin to the previous case, the spin-phonon interaction for ensemble $b$ is rotated out in such interaction picture due to fact that phonon annihilation and creation operator would accumulate the fast oscillating phase factor as determined by $e^{-if_rt}$ and  $e^{+if_rt}$ respectively. Similar to the previous TMS case, the phonon annihilation and creation operators in the local displacement caused by the imbalance in polarizations of the Raman beams, also accumulate the fast oscillating phase factors $e^{-if_rt}$ and  $e^{+if_rt}$. Therefore the terms proportional to $\epsilon_{l}$ are rotated out.  

\subsubsection{Adiabatic elimination of the phonon mode to obtain effective spin-spin interaction between a and c ensembles}
Following the same procedure as that of the entangling operation stated before, we consider the COM  phonon to be fast oscillating on the time scale of $(\Delta_{M}^{ac})^{-1}$ compared to the time scale of the internal spin dynamics i.e. $\{g_{a}/\sqrt{N_a},g_{c}/\sqrt{N_c}\}<4\Delta_{M}^{ac}$. In this case, the  phonon dynamics can be averaged out to zero on the time scale of the spin dynamics and the phonon mode adiabatically follows the spin. We thus adiabatically eliminate the phonon mode to realize an effective spin-spin interaction of the form:
\begin{align}
\hat{H}_{ac}=-\frac{1}{4\Delta_{M}^{ac}} \pc{\mathcal{G}_{a} \mathcal{\hat{S}}_{+}^{a}+ \mathcal{G}_{c}\mathcal{\hat{S}}_{+}^{c}} \pc{\mathcal{G}_{a} \mathcal{\hat{S}}_{-}^{a}+ \mathcal{G}_{c}\mathcal{\hat{S}}_{-}^{c}}+\delta_{ac}\pc{\mathcal{\hat{S}}_{z}^{a}+\mathcal{\hat{S}}_{z}^{c}}.
\label{eq:effenuclear_spin_spinHam}
\end{align}

\subsubsection{Schwinger-Boson representation and effective beam-splitter (BS) operation between  a and c ensembles}
 
We now write the collective spin operator in terms of the Schwinger-Boson representation such that for $l=\{a,c\}$, we have 
\begin{align}
\mathcal{\hat{S}}_{+}^{l}=\hat{l}^{\dagger}_{\uparrow}\hat{l}_{\downarrow}, \quad  \mathcal{\hat{S}}_{+}^{l}=\hat{l}^{\dagger}_{\uparrow}\hat{l}_{\downarrow},\quad
\mathcal{\hat{S}}_{z}^{l}=\frac{1}{2}\pc{\hat{l}^{\dagger}_{\uparrow}\hat{l}_{\uparrow}-\hat{l}^{\dagger}_{\downarrow}\hat{l}_{\downarrow}}, 
\end{align}
For this stage of the protocol, we consider that the ensembles $a$ and $c$ are spin-polarized along the same initial direction $-x$. This is a good approximation if the population in ensemble $a$ in the initial direction $-x$ doesn't appreciably change during the first stage of the protocol. Moreover, we assume that the spin polarized state of ensemble $c$, which we want to teleport,  has slight yet unknown deviation form direction $-x$ direction. The mean field dynamic with such initial conditions reflects that the states associated to the boson $\hat{l}_{\downarrow} $, $\hat{l}^{\dagger}_{\downarrow} $ are macroscopically populated to a good approximation in the short time limit. We therefore replace the operators with their mean values as  $\hat{l}_{\downarrow}=\sqrt{N_{l}}$. This allows us to rewrite the Eq.~\eqref{eq:effenuclear_spin_spinHam} as
\begin{align}
\hat{H}_{ac}=-\frac{1}{4\Delta_{M}^{ac}}\pr{N_{a}\mathcal{G}_{a}^{2}\hat{a}^{\dagger}_{\uparrow}\hat{a}_{\uparrow}+N_{c}\mathcal{G}_{c}^{2}\hat{c}^{\dagger}_{\uparrow}\hat{c}_{\uparrow}}+\delta_{ac}\pr{\hat{a}^{\dagger}_{\uparrow}\hat{a}_{\uparrow}+\hat{c}^{\dagger}_{\uparrow}\hat{c}_{\uparrow}}-\frac{\mathcal{G}_{a}\mathcal{G}_{c}\sqrt{N_{a}N_{c}}}{4\Delta_{M}^{ac}}\pr{\hat{a}^{\dagger}_{\uparrow}\hat{c}_{\uparrow}+\hat{c}_{\uparrow}^{\dagger}\hat{a}_{\uparrow}}
\label{eq:nuclear_spin_Bosonic}
\end{align}
In order to cancel the unwanted terms in the interaction, we set 
\begin{align}\label{eqApp:BeamsplitterUnwantedCondietion21}
(N_a \mathcal{G}_{a}^2)/(4\Delta_{M}^{ac})=(N_{c}\mathcal{G}_{c}^2)/(4\Delta_{M}^{ac})=\delta_{ac}.
\end{align} Eq.~\eqref{eqApp:BeamsplitterUnwantedCondietion21} allows us to calculate the interaction picture transformation factor $f_{r}$ in terms of the system parameters. To do so, we assume that $N_{a}=N_{c}\equiv\bar{N}$ and $\mathcal{G}_{a}= \mathcal{G}_{c}\equiv \mathcal{G}$, then we have $(\bar{N}\mathcal{G}^2)/(4\Delta_{M}^{ac})=\delta_{ac}$. This is further expanded in terms of the factor $f_{r}$ as
\begin{align}\label{eqApp:BeamsplitterUnwantedCondietion}
(\bar{N} \mathcal{G}^2)=4(\Omega-f_{r})(\delta_{M}-f_{r}).
\end{align}
This Equation is solved for $f_{r}$ to get $\textstyle{f_{r}=(1/2)\pc{\delta_{M}+\Omega+\sqrt{g_a^2+\delta_{M}^{2}-2\delta_{M}\Omega+\Omega^{2}} } }$. It is ensured that parameters used in the numerical results allows $f_{r}$ to readily satisfy the time scale given by $\{\delta_{ac},\mathcal{G}_{a}\sqrt{N_{a}},\mathcal{G}_{b}\sqrt{N_{b}},\mathcal{G}_{c}\sqrt{N_{c}},\Delta_{M}^{ac}\}\ll f_{r}$. By this choice of parameters, we thus get the final form of the Hamiltonian resulting in a beam-splitter (BS) operation as given by
\begin{align}
\hat{H}_{\rm BS}=&-\chi_{ac}\bar{N} \pr{\hat{a}^{\dagger}\hat{c}+\hat{c}^{\dagger}\hat{a}},
\label{eq:nuclear_spin_Bosonic form}
\end{align}
where we have omitted the redundant labelling of $\uparrow, \downarrow$.
Under the effective beam-splitter interaction, the quadrature is mixed according to the equations:
\begin{align}
&\hat{X}_{a}=\sin{\theta}\hat{P}_{c}+\cos{\theta}\hat{X}_{a},\nonumber\\
&\hat{P}_{a}=-\sin{\theta}\hat{X}_{c} +\cos{\theta}\hat{P}_{a},\nonumber\\
&\hat{X}_{c}=\cos{\theta}\hat{X}_{c}+\sin{\theta}\hat{P}_{a},\nonumber\\
&\hat{P}_{c}=\cos{\theta}\hat{P}_{c} -\sin{\theta}\hat{X}_{a},
\end{align}
where $\theta= \bar{N}\chi_{ac} t$. We truncate this interaction at times $t=t_{\rm BS}$ such that $\theta=\pi/4$. Therefore, in terms of total interaction time, we obtain $\hat{X}_{a} (t_f)=  (\hat{P}_{c}(0) + \hat{X}_{a}(t_{\rm TMS}))/\sqrt{2}$ and $\hat{X}_{c}(t_f)= (\hat{X}_{c}(0)+\hat{P}_{a}(t_{\rm TMS}))/\sqrt{2}$, as stated in the main text.
%

\section{Entangling Dynamics  and average teleportation Fidelities in the Holstein-Primakoff approximation}\label{Sup:Section3}

The two-mode squeezing Hamiltonian in Eq.~\eqref{eq:SuppTMS} give rise to equation of motion for the vector of spin excitation variables $\textstyle {\mathbf{\hat{u}}=[\hat{a},\hat{a}^{\dagger},\hat{b},\hat{b}^{\dagger}, \hat{c},\hat{c}^{\dagger}}]^T$ as given by
\begin{align}
\partial_{t}\mathbf{\hat{u}} = \mathbf{\mathbf{k}}~\mathbf{\hat{u}}.
\end{align}
Here $\mathbf{\mathbf{k}}$ is a $6\times6$ kernel matrix with its non-zero matrix elements taking the form
\begin{align}
&k_{11}=-i(\delta_{ab}- N_{a}\chi_{a}), ~~ k_{14}=i\chi_{ab}\sqrt{N_{a}N_{b}},\nonumber\\
&k_{22}=i(\delta_{ab}- N_{a}\chi_{a}),~~ k_{23}=-i \chi_{ab}\sqrt{N_{a}N_{b}} ,\nonumber\\
&k_{32}= i \chi_{ab}\sqrt{N_{a}N_{b}} ,~~ k_{33}=-i(\delta_{ab}- N_{b}\chi_{b}) ,\nonumber\\
&k_{41}= -i \chi_{ab}\sqrt{N_{a}N_{b}},~~ k_{44}= i(\delta_{ab}- N_{b}\chi_{b}).
\end{align}

This set of coupled equations allows us to construct the dynamics of the covariance matrix elements $C_{ij}\equiv (1/2) \langle \hat{u}_{i} \hat{u}_{j}+ \hat{u}_{j} \hat{u}_{i}\rangle $, where $\hat{u}_{i}$ and $\hat{u}_{j}$ are the components of vector $\mathbf{\hat{u}}$. The covariance matrix follows the dynamical equation of motion:
\begin{align}
\dot{\mathbf{\mathbf{C}}}(t) = \mathbf{k} \mathbf{C}(t) + \mathbf{C}(t)  \mathbf{k}^{T}.
\label{eq:CovDynaEOM}
\end{align} 
The covariance matrix in terms of the position and momentum quadrature of the system is given by $\mathbf{C}_{XP}(t)=\mathbf{R}{\mathbf{C}}(t) \mathbf{R}^T$, where the transformation $\mathbf{R}$ is a block matrix:
\begin{align}
\mathbf{R} = 
\left(\begin{array}{ccc} \mathbf{T} & \mathbf{0}   &  \mathbf{0} \\ \mathbf{0}&  \mathbf{T} &
\mathbf{0}\\
\mathbf{0} & \mathbf{0}&   \mathbf{T} \end{array} \right),\quad {\rm where} \quad 
\mathbf{T}=\left(
\begin{array}{cc}
 \frac{1}{\sqrt{2}} & \frac{1}{\sqrt{2}}  \\
 -\frac{i}{\sqrt{2}} & \frac{i}{\sqrt{2}}
\end{array}
\right).
\end{align}

The $6\times 6$ covariance matrix $\mathbf{C}_{XP}(t)$ capture fully the dynamics of the system as dictated by the bi-linear form of the Hamiltonian in Eq. \eqref{eq:nuclear_spin_Bosonic} and given by

\begin{align}
\mathbf{\mathbf{C}}_{XP}(t) = 
\left(\begin{array}{@{}cc|c@{}} \mathbf{\mathbf{C}}_{a a } & \mathbf{\mathbf{C}}_{ab}   &  \mathbf{\mathbf{C}}_{ac} \\  \mathbf{\mathbf{C}}_{ab}^{T} & \mathbf{\mathbf{C}}_{b b} &
\mathbf{\mathbf{C}}_{bc}\\ \hline
\mathbf{\mathbf{C}}_{ac}^{T} & \mathbf{\mathbf{C}}_{bc}^{T} &  \mathbf{\mathbf{C}}_{c c}  \end{array} \right),
\end{align}
where $\mathbf{\mathbf{C}}_{a a}, \mathbf{\mathbf{C}}_{bb} $ and $\mathbf{\mathbf{C}}_{cc}$ are the local  covariance  matrices associated to the spin ensembles $a$,  $b$, and $c$, respectively.
Moreover the matrix $\mathbf{\mathbf{C}}_{ab}$ captures the correlation for $a$ and $b$ mode. 
Here $[.]^{T}$ represents the matrix transpose operation. The $4\times 4$ partition matrix $\mathbf{\mathbf{C}}_{AB}$ represents the entangling dynamics of the $a$ and $b$ ensembles for $\chi_{aa}\bar{N}=\chi_{ba}\bar{N}\simeq\delta_{ab} $. It closely follows the  solution of the form
\begin{align}
\mathbf{\mathbf{C}}^{AB}(t)\simeq 
\left(\begin{array}{cccc} 
(1/2)\cosh{2r} & 0  &  0 &(1/2)\sinh{2r}\\   
0 & (1/2)\cosh{2r}  &  (1/2)\sinh{2r} & 0\\ 
0 & (1/2)\sinh{2r}  &  (1/2)\cosh{2r} & 0\\
(1/2)\sinh{2r} & 0  &  0 & (1/2) \sinh{2r}
\end{array} \right),
\end{align}
where $r=\bar{N}\chi_{ab} t $ with $\chi_{ab}<0$. It has been shown that for coherent states \cite {Masashi2004} (i.e. spin coherent state in the HP approximation), the average teleportation fidelity  is determined by the evolution of these matrix elements and given by
\begin{align}
\bar{F}_{\rm SC}^{\rm HP}\simeq\frac{1}{1+2 \mathbf{C}^{AB}_{11}+\mathbf{C}^{AB}_{14}+\mathbf{C}^{AB * }_{23}}= \frac{1}{1+e^{2 r}}.
\end{align}
In the limit $r\rightarrow 0$, we recover the classical limit $\bar{F}_{\rm SC}^{\rm HP}\rightarrow 1/2$ \cite{Hammerer2005}. 
We plot this expression in Fig.~\ref{fig:telepfig4} of the main text and compare this analytical solution in the HP picture with the exact numerical simulation of the teleportation protocol. 
For spin-squeezed states, which maps to a bosonic squeezed vacuum state in the HP picture,  we follow the analytical expression  of the teleportation fidelity given by \cite{SSAVTFed} 
\begin{align}
\bar{F}_{\rm SS}^{\rm HP}\simeq\frac{1}{\sqrt{(1+e^{2 r} (\xi_{c}^{in})^2)(1+\pr{e^{2 r}/(\xi_{c}^{in})^2})}},
\end{align}
where $\xi_{c}^{in}$ is the amount of input squeezing in  the $c$ ensemble. We  obtained its value under one axis-twisting dynamics when studied in the HP picture.
\section{Mean field equation of motion for the full model}\label{Sup:Section4}
The lab frame Hamiltonian of the full system is given by
\begin{align}\label{eqAPP:FullLabFrameHamlt}
\hat{H}=&\sum_{l=a,b,c}\Omega_{l}\hat{S}_{x}^{l}+\sum_{l=a,b,c}\mathcal{G}_{l}(\hat{m}+\hat{m}^{\dagger})\hat{S}_{z}^{l}+\delta_{M}\hat{m}^{\dagger}\hat{m}. 
\end{align}
The corresponding mean field equations for the four-partite system are obtained by using the Hamiltonian in Eq.~ \eqref{eqAPP:FullLabFrameHamlt} and given by

\begin{align}
 &\partial_{t}\av{\hat{S}_{x}^{l}}=-\mathcal{G}_{l}\av{\hat{m}}\av{\hat{S}_{y}^l}-\mathcal{G}_{l}\av{\hat{m}^{\dagger}}\av{\hat{S}_{y}^l}\\
 &\partial_{t}\av{\hat{S}_{y}^{l}}=\mathcal{G}_{l}\av{\hat{m}}\av{\hat{S}_{x}^l}+\mathcal{G}_{l}\av{\hat{m}^{\dagger}}\av{\hat{S}_{x}^l}-\Omega_{l}\av{\hat{S}_{z}^l}\\
 &\partial_{t}\av{\hat{S}_{z}^{l}}=\Omega_{l}\av{\hat{S}_{y}^l},\quad\text{where }l=a,b,c\\
   &\partial_{t}\av{\hat{m}}=-i\mathcal{G}_{a}\av{\hat{S}_{z}^a}-i\mathcal{G}_{b}\av{\hat{S}_{z}^b}-i\mathcal{G}_{c}\av{\hat{S}_{z}^c}-i\delta_{M}\av{\hat{m}}\\
 &\partial_{t}\av{\hat{m}^{\dagger}}=i\mathcal{G}_{a}\av{\hat{S}_{z}^a}+i\mathcal{G}_{b}\av{\hat{S}_{z}^b}+i\mathcal{G}_{c}\av{\hat{S}_{z}^c}+i\delta_{M}\av{\hat{m}^{\dagger}}
 \end{align}

We simulate the full model dynamics by performing numerical simulations of the above equation using the discrete truncated Wigner approximation (DTWA) \cite{Blakie2008,Polkovnikov2010SM,Schachenmayer2015SM}. This method accounts for  the quantum dynamics by averaging over an ensembles of classical trajectories by  sampling the initial conditions such that we recover the  correlations   of the initial state. The DTWA reproduces    the  quantum  dynamics of one and two point observables, appropriately incorporating  beyond mean-field effects.

\section{ Final local Rotations on ensemble b}\label{Sup:Section5}
In this section, we derive the form of rotation  that we apply on ensemble $b$ to complete the teleportation protocol. As stated in the main text, the amount of the rotation depends on the two classically communicated measurement outcomes $\beta_{z}$ and $\beta_{y}$ that are associated with the measured valued of the operator $\hat{S}_{z}^{a,c}$, as performed on ensembles $a$ and $c$. We extract the rotation from the underlying HP picture of the corresponding $b$ bosonic mode. Let us consider the general form the displacement operator for the $b$ ensemble bosonic mode viz:
\begin{align}
\label{eq:SuppDisplref}
\hat{D}(\beta)=\exp(\beta\hat{b}^{\dagger}-\beta^*\hat{b})=\exp(i\sqrt{2}(\beta_i\hat{X}_b-\beta_r\hat{P}_b))\simeq\exp\pr{i\sqrt{2}\pr{\beta_i(-\sqrt{(2/N_b)}\hat{S}_z^b)-\beta_r(\sqrt{(2/N_b)}\hat{S}_y^b)}}, 
\end{align}
where we have used spin-to-boson mapping $\hat{P}_b(t)=\sqrt{(2/N_b)}\hat{S}_y^b(t)$ and $\hat{X}_b(t)=-\sqrt{(2/N_b)}\hat{S}_z^b(t)$ as stated in the main text. In addition, from the main text we have:
\begin{align}
\label{eq:SMMeasureACref}
&\hat{S}_{y}^{{\rm ideal},b}(t_f)= \sqrt{2} \hat{S}_{z}^{ a}(t_f) -\hat{S}_{y}^{c}(0)\\
&\hat{S}_{z}^{{\rm ideal},b}(t_f)=- \sqrt{2} \hat{S}_{z}^{ c}(t_f) +\hat{S}_{z}^{c} (0).
\end{align}
To map the input state of ensemble $c$ to the output state of $b$, we first remove the displacement $\sqrt{2} \hat{S}_{z}^{ a}(t_f)$ and $- \sqrt{2} \hat{S}_{z}^{ c}(t_f)$ from $\hat{S}_{y}^{{\rm ideal},b}(t_f)$ and $\hat{S}_{z}^{{\rm ideal},b}(t_f)$ in these equations. To do so,  we consider from the above Eqs. that $\hat{S}_{y}^{{\rm ideal},b}(t_f)= \sqrt{2} \hat{S}_{z}^{ a}(t_f)$ and $\hat{S}_{z}^{{\rm ideal},b}(t_f)=- \sqrt{2} \hat{S}_{z}^{ c}(t_f)$. Using these expressions and spin-to-boson mappings for ensemble $b$, we write 
\begin{align}
 \label{eq:supphatphatxfinal1}
    \hat{P}_b(t_f)=(2/\sqrt{N_b})[\hat{S}_{z}^{ a}(t_f)],
    \end{align}
    \begin{align}
    \hat{X}_b(t_f)=(2/\sqrt{N_b})[\hat{S}_{z}^{ c}(t_f)]. 
    \label{eq:supphatphatxfinal2}
\end{align}
Now for the operator $\hat{b}=(1/\sqrt{2})(\hat{X}_b+i\hat{P}_b)\equiv\hat{\beta}_r+i\hat{\beta}_i$, the real and imaginary part of its complex eigenvalues are $\beta_{r}\sim\langle\beta|\hat{X}_{b}|\beta\rangle$, $\beta_{i}\sim\langle\beta|\hat{P}_{b}|\beta\rangle$. These are respectively the position and momentum coordinates of the bosonic phase space distribution. Using the expression in Eqs.~\eqref{eq:supphatphatxfinal1} and \eqref{eq:supphatphatxfinal2}, we obtain $\hat{\beta}_r=\sqrt{(2/N_b)}\hat{S}_{z}^{ c}(t_f)$ and $\hat{\beta}_i=\sqrt{(2/N_b)}\hat{S}_{z}^{ a}(t_f)$. Therefore, for each of the measured values of $\hat{S}_{z}^{c,a}(t_f)$, the projected state of ensemble $b$ is displaced in the phase space by an amount determined by the eigenvalues $\beta_{r}=\sqrt{(2/N_b)}M_{z}^{c}(t_f), \beta_{i}=\sqrt{(2/N_b)}M_{z}^{ a}(t_f)$, which are the momentum and position coordinates, respectively. On the corresponding Bloch sphere, the projected state is rotated by the same amounts along the axis $\hat{S}_{z}^{b}$ and $\hat{S}_{y}^{b}$, respectively. To undo this rotation, we apply $\hat{D}^{\dagger}(\beta)=\hat{D}(-\beta)$ as obtained from Eq.~\eqref{eq:SuppDisplref}. Inserting the eigenvalues $\beta_{r},\beta_{i}$ in the expression of $\hat{D}(-\beta)$ and by simplifying, we obtain
\begin{align}
\hat{D}(-\beta)\simeq\exp\pr{i\frac{2\sqrt{2}}{N_b}\pr{M_{z}^{a}(t_f)\hat{S}_z^b+M_{z}^{c}(t_f)\hat{S}_y^b}}=\exp\pr{i(2/N_b)\pc{\beta_{y}\hat{S}_y^b+\beta_{z}\hat{S}_z^b}},
\end{align}
where we have defined $\beta_{y}\equiv \sqrt{2} M_{z}^{c}(t_f) $ and  $\beta_{z}\equiv \sqrt{2} M_{z}^{ a}(t_f)$. We call $\hat{D}_{r}({\beta_{y}, \beta_{z}})\equiv\hat{D}(-\beta)$. This is the amount of rotation that is stated in the main text and it is applied on ensamble $b$ upon receiving the classically communicated message. After performing this rotation, the state of ensemble $b$ becomes
\begin{align}
    &\hat{S}_{y}^{b}(t_f)= -\hat{S}_{y}^{c}(0)\\
    &\hat{S}_{z}^{b}(t_f)=  \hat{S}_{z}^{c} (0), 
\end{align}
which is different then the input state by a sign in the $y$-component. To make the output state (i.e. obtained at $t=t_{f}$) of $b$ same as that of the input state (i.e. prepared at the initial time, $t=0$) of $c$, the ensemble $b$ is subjected to a $\pi$ rotation around the $z$-axis given by $\hat{D}_{\pi}^b=\exp\left(i\pi\hat{S}_z^b\right)$. This rotation corrects the $x$-component of the ensemble $b$, since according to our initial conditions, the state of ensemble $b$ was opposite to that of the ensemble $c$. We thus obtain:

\begin{align}
    &\hat{S}_{y}^{b}(t_f)= \hat{S}_{y}^{c}(0)\\
    &\hat{S}_{z}^{b}(t_f)=  \hat{S}_{z}^{c} (0).
\end{align}

\section{Supporting Numerical Results}\label{Sup:Section6}
\begin{figure*}[h!]
	\includegraphics[width=0.245\columnwidth]{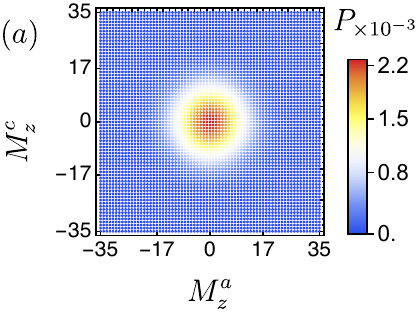}
 \includegraphics[width=0.245\columnwidth]{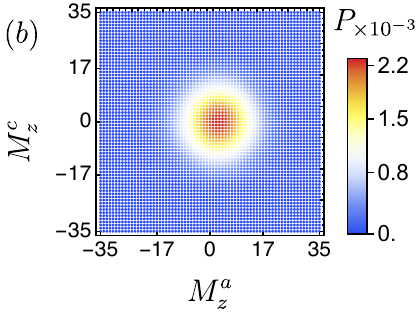}
	\includegraphics[width=0.245\columnwidth]{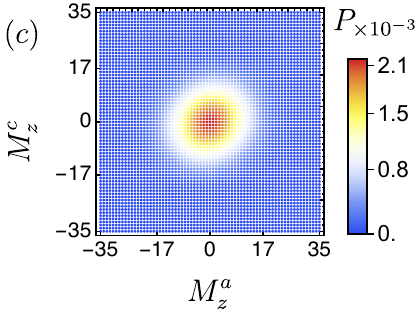}
 	\includegraphics[width=0.245\columnwidth]{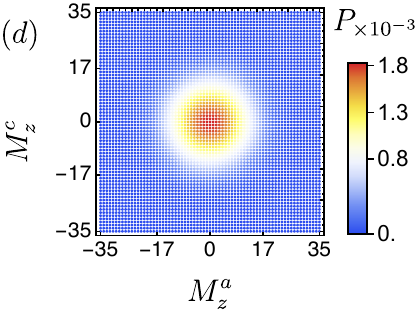}
		\includegraphics[width=0.245\columnwidth]{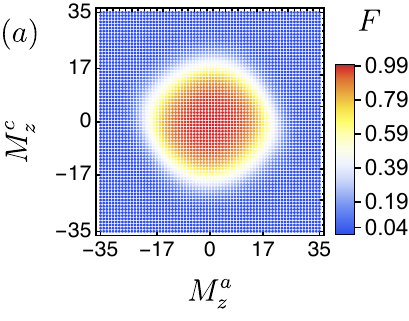}
  \includegraphics[width=0.245\columnwidth]{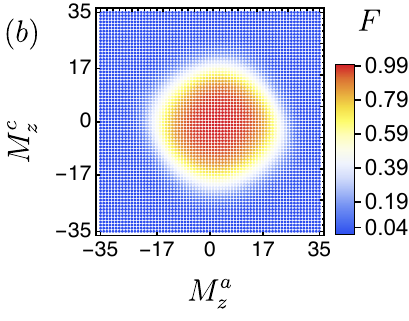}
		\includegraphics[width=0.245\columnwidth]{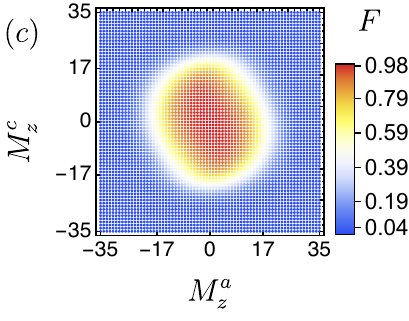}
  		\includegraphics[width=0.245\columnwidth]{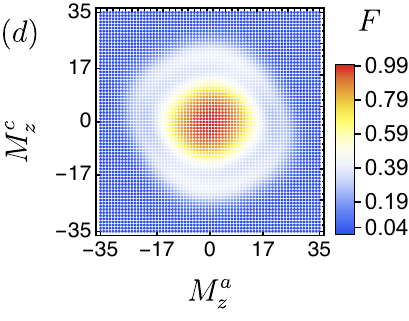}
	\caption{ (Top Row) Population measurement outcome probability distribution for the cases when the input state is: (a-d)  spin-coherent (SC), phase-displaced spin coherent (PDSC), spin squeezed (SS), and Dicke state (DS), respectively. The bottom row represents the corresponding teleportation fidelity distributions, respectively.}
	\label{Appfig:telepfig2}
\end{figure*}
\begin{figure*}[t]
	\includegraphics[width=0.32\columnwidth]{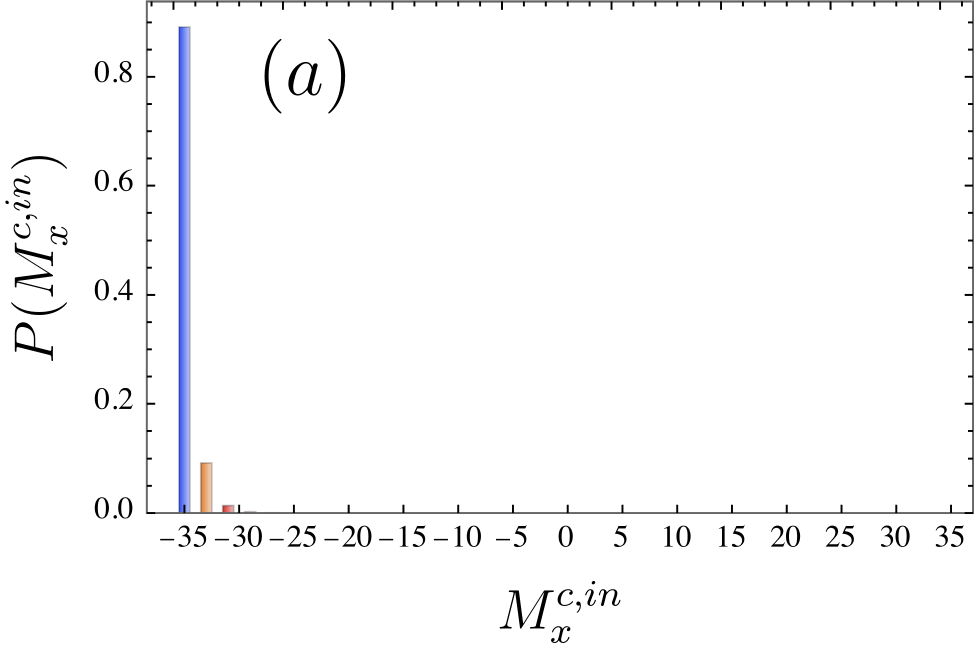}
	\includegraphics[width=0.32\columnwidth]{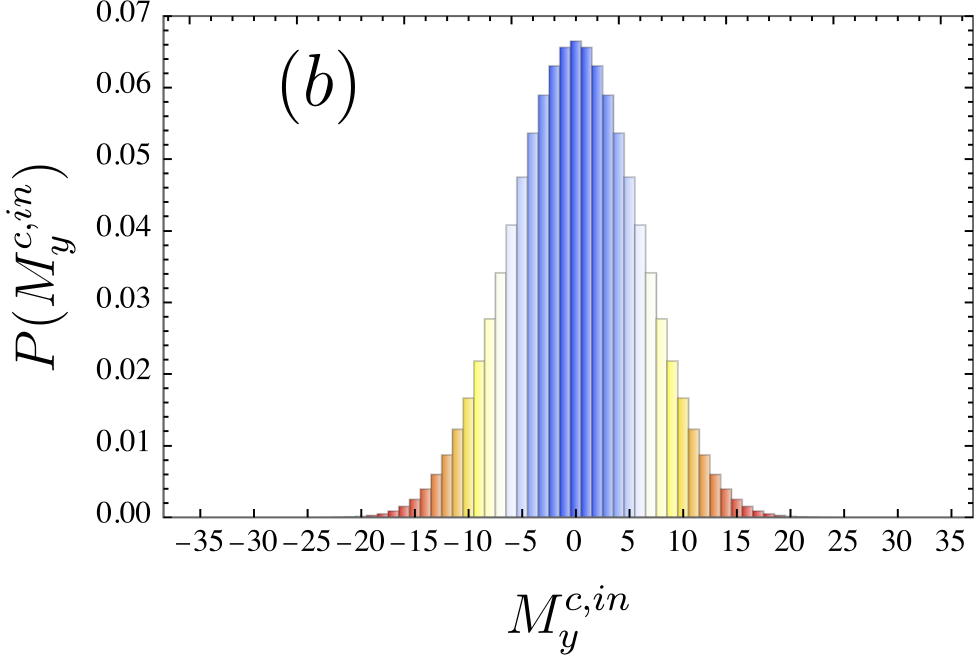}
	\includegraphics[width=0.32\columnwidth]{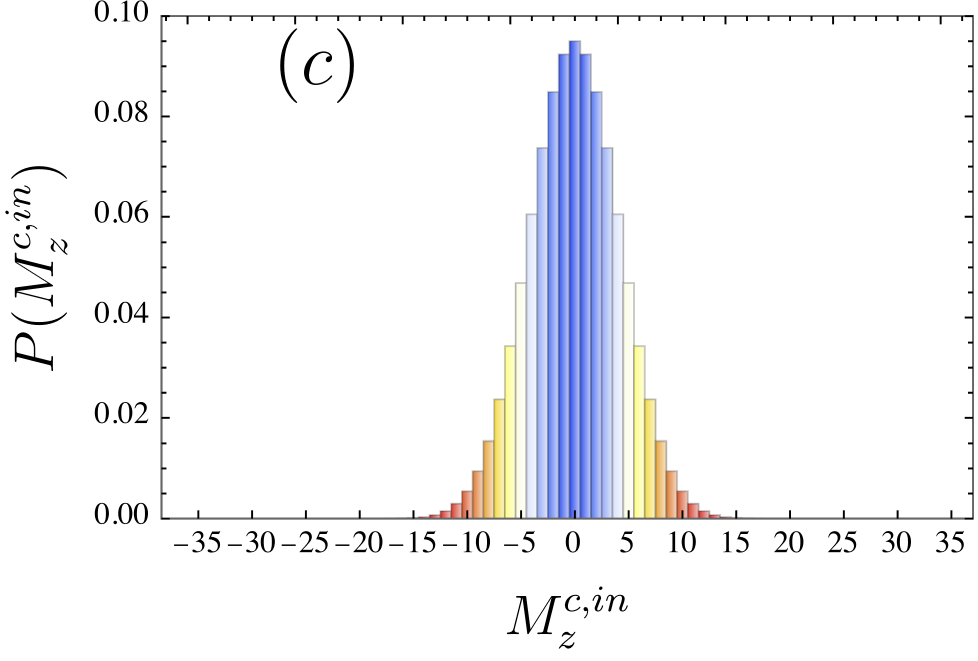}

		\includegraphics[width=0.32\columnwidth]{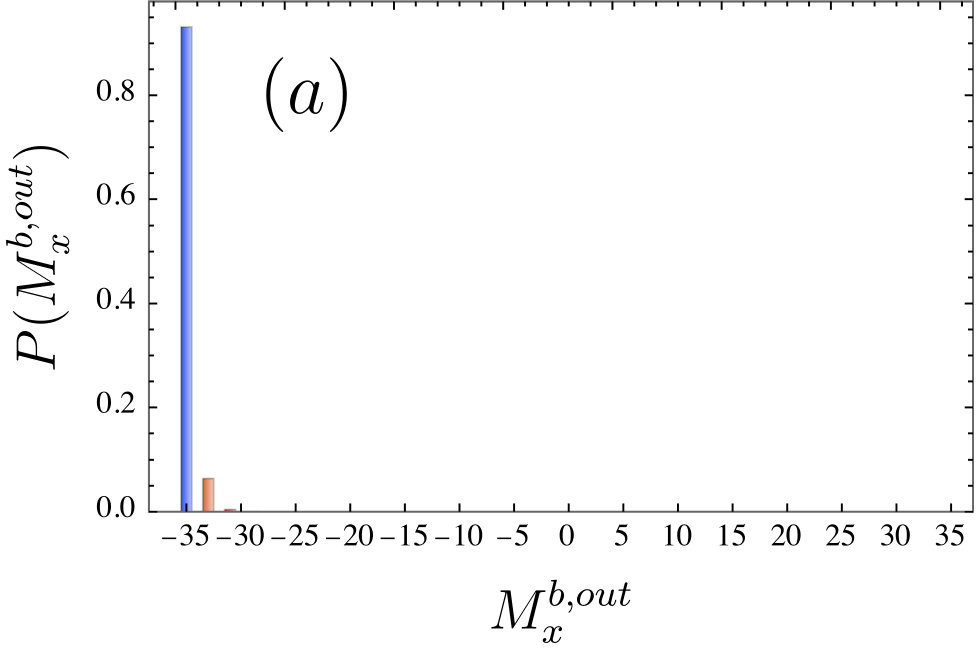}
		\includegraphics[width=0.32\columnwidth]{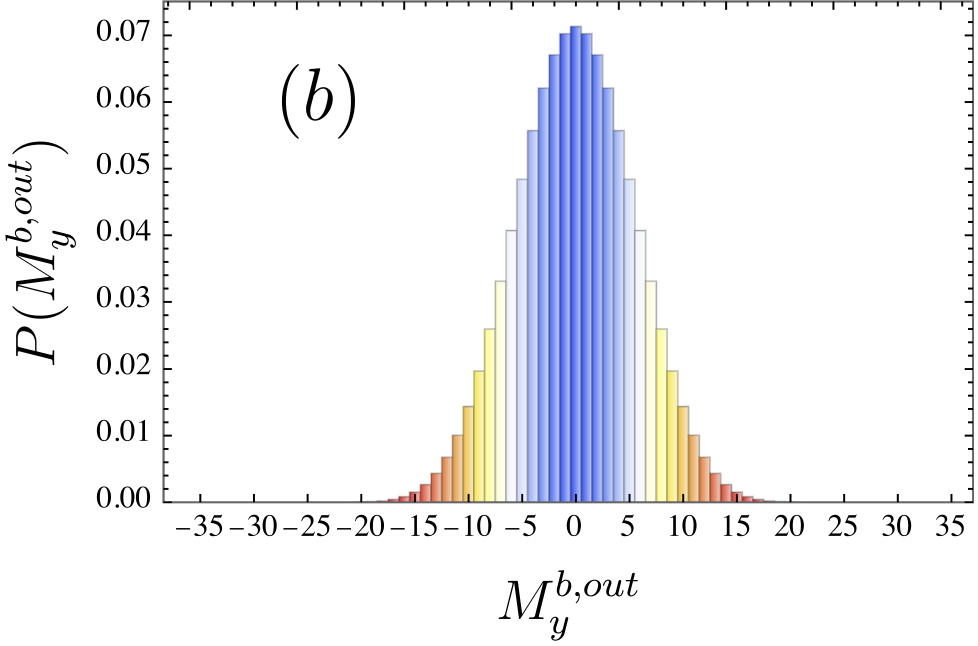}
		\includegraphics[width=0.32\columnwidth]{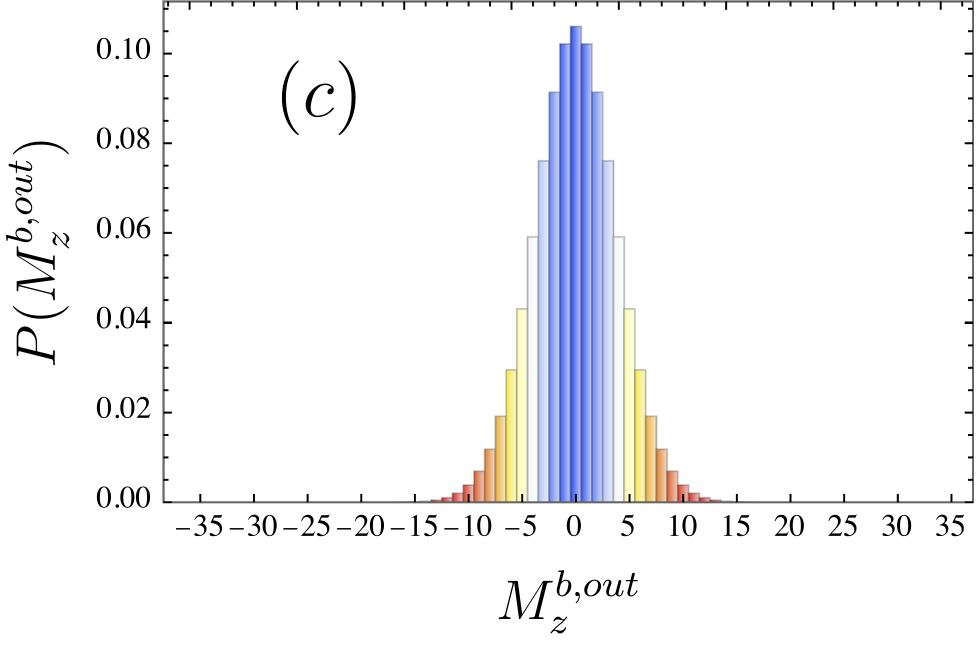}
	\caption{The probability distribution functions of the magnetization  $M_{x},M_{y}, M_{z}$  for the case when input state  is a squeezed state (SS) as described in the main text. The top row correspond to the input state, while the bottom row represents the magnetization of the teleported state for the most probable outcome. The similarity of the statistics ensures  teleportation of the  spin squeezed state.}
	\label{Appfig:telepfig2b}
\end{figure*}
\begin{figure*}[t]
	\includegraphics[width=1\columnwidth]{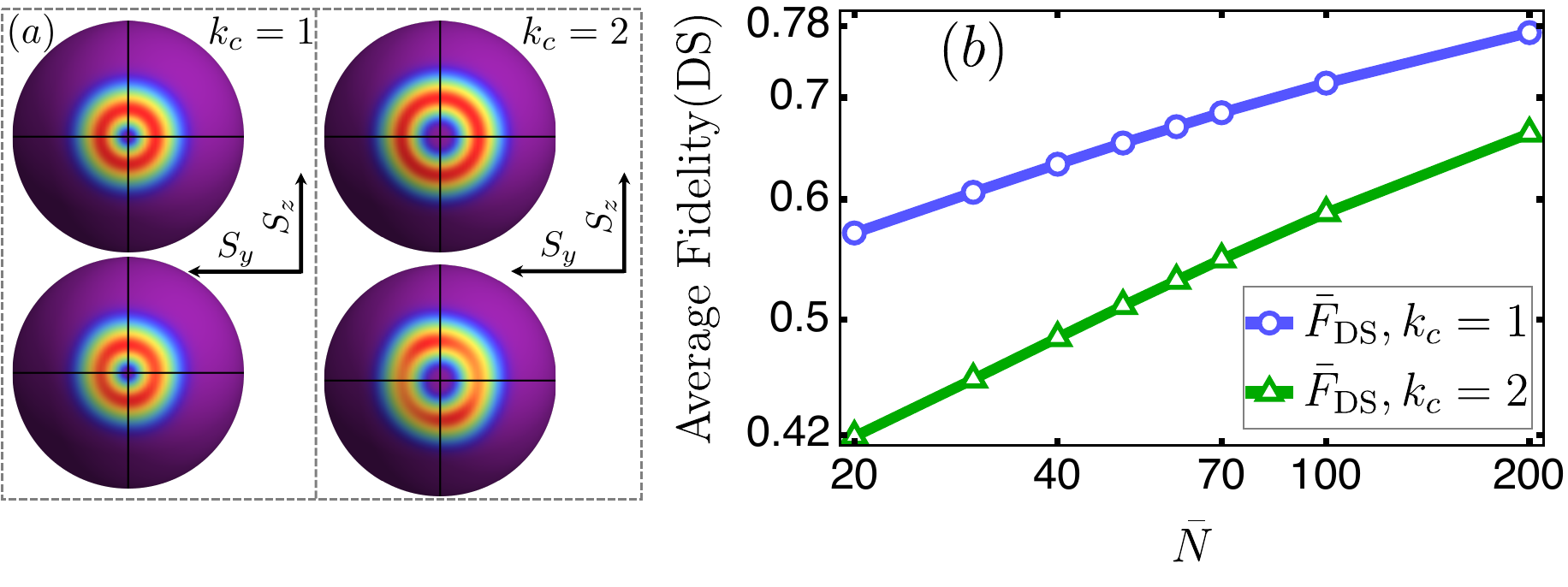}
	\caption{(a) Husimi-Q function for input Dicke states (top row) for spin excitation $k_{c}=1$  and $k_{c}=2$, respectively. Bottom row corresponds to their teleported versions for the most probable outcome obtained for $\bar{N}=70$ ions. (b) We show the corresponding average teleportation fidelity against the number of ions for both cases. The average fidelity increases  monotonically for the cases considered.}
	\label{Appfig:telepfig2c}
\end{figure*}
In this section, we present numerical results that demonstrate  one can  successfully teleport various input states as discussed in the main text and add results for teleporting two-excitation Dicke states.  

In Fig.~\ref{Appfig:telepfig2} (top row), we present the probability distribution functions associated with the measurement outcome as performed on ensemble $a$ and $c$ and discussed in the main text. While in the main text, we only show such a function for a SC state (cf. Fig.~\ref{fig:telepfig3}(a)), here we show distribution functions for all four input states of SC, PDSC, SS and DS. Similarly, in the Fig.~\ref{Appfig:telepfig2} (bottom row), we show the corresponding fidelity distribution functions for different measurement outcomes, respectively. These results allow us  to assess the probability of measuring a fixed outcome with the associated fidelity of the teleported state for that outcome. In general, the teleportation requires averaging over all the possible outcome and an average teleportation fidelity is assessed.

In the main text, we showed that an entangled spin squeezed (SS) state can be teleported under the proposed protocol. For this case, we further numerically compute the probability distributions of the magnetization in different directions $x,y,z$, both for the input $P(M_{x}^{c,in}),P(M_{y}^{c,in}), P(M_{z}^{c,in})$ and teleported state $P(M_{x}^{b,out}),P(M_{y}^{b, out}), P(M_{z}^{b,out})$. These are shown in  the top and bottom rows  of Fig.~\ref{Appfig:telepfig2b} respectively. From (a-c), we observe that spin magnetization statistics along all directions of the input SS state and its teleported version are similar. In particular, the magnetization statistics along $y$ has larger variance in the expense that its variance along $z$ becomes small, which indicates a SS state. This feature is also reflected in its teleported version. 
Additionally, for a SS state, the magnetization $M_x$ always satisfies the property that $M_x+N_{c}/2$ is even.
Such a feature is also observed  in the teleported version of the SS state. 

Finally, we also simulate the teleportation protocol for the case when the input state is a Dicke state with two spin excitation, i.e. with $k_{c}=2$. Preparing such initial Dicke states would require heralding or higher order non-linear interaction (e.g. $\hat S_z \hat S_z$), which are  accessible in the Penning trap geometry but has not been reported yet. Nevertheless, we  study the  $k_{c}=2$ case to see if the present teleportation scheme is successful when input state has more non-classical features. As shown in the Fig.~\ref{Appfig:telepfig2c} (a) (second column), the telepotaion of $k_{c}=2$ state is also possible as witnessed by the similarity in the annular noise distributions of the Husimi-Q functions for input and teleported states. However, when we  compare it  to the case $k_c=1$ (as shown in first column), the noise distribution of the teleported state for $k_c=2$ has lesser radial symmetry. This is due to the fact the our teleportation scheme  relays on the validity of the Holstein-Primakoff approximation  to leading order in $1/N$ which weakens as we increase the number of excitation in the input state. The finite  number of ions, also limit the achievable   EPR correlations in the entangled states needed to resolve the structures in the Husimi Q-function that become  more pronounced in  Dicke state with higher excitation. In Fig.~\ref{Appfig:telepfig2c} (b), we plot the average fidelity against the number of ions $\bar{N}$. Both of the input states with $k_c=1,~ k_{c}=2$ have average fidelity that monotonically increases with $\bar{N}$. 

\bibliographystyle{apsrev4-1}

\end{document}